\documentclass[sigconf,review=false,balance=false, anonymous=false]{acmart}
\usepackage{popets}

\usepackage{amsmath,amssymb,amsfonts}
\usepackage{graphicx}
\usepackage{textcomp}
\usepackage{bmpsize}
\usepackage{comment}
\usepackage{enumitem}
\usepackage{amssymb}
\usepackage{pifont}
\newcommand{\cmark}{\ding{51}}%
\newcommand{\xmark}{\ding{55}}%
\usepackage{makecell}

\usepackage{tikz}
 \UseRawInputEncoding

\usepackage{xcolor}

\setcitestyle{numbers}

\usepackage{amsthm}
\makeatletter
\def\els@aparagraph[#1]#2{\elsparagraph[#1]{#2\@addpunct{.}}}
\def\els@bparagraph#1{\elsparagraph*{#1\@addpunct{.}}}
\makeatother

\usepackage{caption} 
\usepackage{float}
\floatstyle{plaintop}
\restylefloat{table}

\usepackage{xcolor}

\usepackage{array}
\newcolumntype{M}[1]{>{\centering\arraybackslash}m{#1}}

\usepackage{afterpage}

\usepackage{comment}

\usepackage{booktabs}

\usepackage{pdfpages}

\usepackage{algorithmicx}
\usepackage[ruled, vlined]{algorithm2e}
\usepackage{float}

\usepackage{scalerel,stackengine}
\stackMath
\newcommand\reallywidehat[1]{%
\savestack{\tmpbox}{\stretchto{%
  \scaleto{%
    \scalerel*[\widthof{\ensuremath{#1}}]{\kern-.6pt\bigwedge\kern-.6pt}%
    {\rule[-\textheight/2]{1ex}{\textheight}}
  }{\textheight}%
}{0.5ex}}%
\stackon[1pt]{#1}{\tmpbox}%
}

\usepackage{color}

\setcopyright{none}
\renewcommand\footnotetextcopyrightpermission[1]{} 
\usepackage{tabularx,booktabs}

\usepackage{fontawesome}
\makeatletter
\def\DeclareUnicodeCharacter#1#2{}
\makeatother

\usepackage{utfsym}

\usepackage{ragged2e}

\usepackage{etoolbox}
\makeatletter
\patchcmd{\@makecaption}
  {\scshape}
  {}
  {}
  {}
\makeatother

\usepackage{mwe}

\usepackage{array}

\usepackage{multirow}
\usepackage{makecell}

\usepackage{pifont}

\usepackage{amsmath}
\usepackage{bm}
\usepackage{amssymb}
\usepackage{amsthm}

\makeatletter
\newcommand{\removelatexerror}{\let\@latex@error\@gobble}
\makeatother

\usepackage{comment}  

\setcopyright{popets}
\copyrightyear{YYYY}

\acmYear{}
\acmVolume{}
\acmNumber{}
\acmDOI{XXXXXXX.XXXXXXX}
\acmISBN{}
\acmConference{}
\begin{document}

\title[SoK: Data Privacy in Virtual Reality]{SoK: Data Privacy in Virtual Reality}


\author{Gonzalo Munilla Garrido}
\authornotemark[1]\thanks{*Corresponding author.}
\affiliation{%
  \institution{TU Munich}
  \city{Munich}
  \country{Germany}}
\email{gonzalo.munilla-garrido@tum.de}

\author{Vivek Nair}
\affiliation{%
  \institution{UC Berkeley}
  \city{Berkeley}
  \country{USA}}
\email{vcn@berkeley.edu}

\author{Dawn Song}
\affiliation{%
  \institution{UC Berkeley}
  \city{Berkeley}
  \country{USA}}
\email{dawnsong@berkeley.edu}


\renewcommand{\shortauthors}{Garrido et al.}

\begin{abstract}
The adoption of virtual reality (VR) technologies has rapidly gained momentum in recent years as companies around the world begin to position the so-called ``metaverse'' as the next major medium for accessing and interacting with the internet. While consumers have become accustomed to a degree of data harvesting on the web, the real-time nature of data sharing in the metaverse indicates that privacy concerns are likely to be even more prevalent in the new ``Web 3.0.'' Research into VR privacy has demonstrated that a plethora of sensitive personal information is observable by various would-be adversaries from just a few minutes of telemetry data. On the other hand, we have yet to see VR parallels for many privacy-preserving tools aimed at mitigating threats on conventional platforms. This paper aims to systematize knowledge on the landscape of VR privacy threats and countermeasures by proposing a comprehensive taxonomy of data attributes, protections, and adversaries based on the study of 74 collected publications. We complement our qualitative discussion with a statistical analysis of the risk associated with various data sources inherent to VR in consideration of the known attacks and defenses. By focusing on highlighting the clear outstanding opportunities, we hope to motivate and guide further research into this increasingly important field.
\end{abstract}


\maketitle



\section{Introduction}
\label{sec:Introduction}

Virtual reality (VR) has recently become a major investing target for leading tech industry players aiming towards the so-called ``metaverse'' \cite{BlackRockOnMeta}, a paradigm shift towards an internet in the form of a 3D virtual world.
This new internet would require VR devices such as headsets and hand-held controllers to digitize and relay users' physical characteristics and movements to other users worldwide for immersive interaction.
While the ``metaverse'' might hold promise, researchers have recently shown how easily an attacker could identify \cite{S30, S34, S55} and profile \cite{S52, S46, S15} VR users with a few minutes of data streaming, and demonstrated that the scope and scale of data collection in VR supersedes the capabilities of current internet platforms \cite{S15}. 
Researchers further illustrated how malicious developers could hide artifacts in virtual environments that inconspicuously induce users to reveal personal information, e.g., by playing a seemingly innocent game \cite{S15}.
These attacks are partly possible due to VR's unparalleled immersiveness, which can make users more susceptible to self-disclosure \cite{S4, S6}, and social engineering \cite{S13, S17}.
 
Unlike the current internet platforms, where users can employ Tor, VPNs, proxies, and incognito mode to fend off user tracking and profiling, there is no equivalent and mature defense suite for combating unique threats in VR.
The extant literature offers a scattered set of privacy defenses at a proof-of-concept stage 
with no significant knowledge transfer to commercial-grade applications.
Industry practices are not encouraging either, as VR devices showed vulnerabilities \cite{S21}, some developers ignore their own privacy policies, and their updates trend towards more data harvesting \cite{S22}.

The looming unprecedented privacy threats in VR entail a ``dysto-pian metaverse'' as long as we lack the robust defenses available on the current internet.
With this study, we aim to begin tackling this impending challenge by providing new, comprehensive threat and defense models, and classifications of data attributes, attacks, and defenses at the intersection of VR and privacy. 
Specifically, our SoK evaluates, systematizes, and contextualizes $74$ studies filtered from over $1700$ publications.
The structure of the SoK flows naturally in the context of privacy in VR: Sensitive data attributes are targeted and harvested by adversaries.
In response, privacy practitioners design the corresponding defenses.
Each taxonomy we introduce builds upon the previous classification, e.g., we classified VR attacks based on our proposed threat model and attributes taxonomy.
Lastly, we outline privacy opportunities by highlighting some of the most vulnerable yet easiest-to-protect attributes in VR. 

With the following \textbf{contributions}, we hope to guide researchers to find the right privacy opportunities quickly:

\begin{enumerate}
    \item We proposed a holistic information flow, threat, and defense models to frame future studies (\S\ref{sec:threat_model}).
    \item We built a taxonomy of data attributes collectible in VR clustered in $9$ classes with more than $60$ data points (\S\ref{sec:data_taxonomy}).
    \item We categorized $34$ attacks (\S\ref{sec:vr_attacks}) and $35$ defenses (\S\ref{sec:vr_defenses}), and focused our insights by answering $10$ research questions.
    \item We performed a quantitative and qualitative analysis to outline practices and opportunities (\S\ref{sec:opportunities}), and extracted $12$ findings and future work items to advance VR defenses (\S\ref{sec:discussion}).
\end{enumerate}

\section{Background \& Related Work}
\label{sec:background_rw}

In the context of the reality and virtuality continuum of Milgram and Kishino \cite{MR_continuum}, our SoK focuses on virtual reality (VR), i.e, a computer-simulated interactive environment experienced in the first person \cite{metaverse_regulation}.
The considered VR devices stem from reviewing the $74$ collected publications.

\subsection{VR Devices}
\label{subsec:vr_devices}

Since the wave of mature VR products of 2016, the wider public has experienced immersive VR like never before.
Primarily, users employ a head-mounted display (HMD) with (integrated) speakers, a microphone, and two handheld controllers with buttons \cite{meta_quest_2}.
Some HMDs tether to a PC \cite{vive_pro}, while others can remain wireless \cite{meta_quest_2}. 
The VR system tracks the HMD and the controllers by outside-in tracking (using stationary external sensors \cite{vive_tracker}) or inside-out tracking (employing built-in optical sensors and inertial measurement units: three-axis accelerometers and gyroscopes \cite{meta_quest_2}).
Front cameras for inside-out tracking also enable the user to observe their real-world surroundings.
This basic setup generates realistic 3D graphics, spatial audio, verbal interaction, and six degrees of head and hand tracking (X, Y, and Z positions, and yaw, pitch and roll)
Manufacturers design VR devices for use in a ``controlled'' environment (e.g., home, backyard, office, etc.).

Other devices make VR experiences more immersive yet more pervasive, as they require additional sensitive user input.
Optical sensors for eye-tracking enable foveated rendering \cite{HTC_vive_pro_eye}, increasing the quality of the visual output~\cite{foveated_Rendering} and lengthening HMD battery life by reducing GPU load.
Moreover, eye-tracking combined with additional optical sensors that register facial features (face trackers \cite{facial_tracker}) enables telepresence via expressive (photorealistic) avatars \cite{Codec_avatars}.
Handheld controllers (and HMDs \cite{PSV2}) can provide haptic feedback, have touch sensors for detecting holding gestures \cite{Vive_Cosmos}, and the latest models have outward cameras for improved tracking \cite{meta_quest_pro}.
A natural transition is forced feedback gloves that provide more ergonomic and realistic interactions \cite{haptic_gloves}; in contrast, users can also employ conventional keyboards \cite{S68}.
Full-body tracking \cite{full_body} enables more expressive and richer experiences with other users in virtual worlds.
Additionally, users may dawn haptic vests \cite{haptic_vest} that deliver positional haptic feedback and prototypical masks that emulate smells \cite{VR_smell}.
Healthcare VR applications include sensors that measure galvanic skin response \cite{S64}, electrodermal activity \cite{Amelia_Virtual_Care}, heart rate \cite{S57}, skin temperature \cite{S61} and measure brain waves (HMDs with EEGs for brain-computer interfacing) \cite{EEG_HMD}

This plethora of sensors and feedback devices facilitate immersive digital interactions in VR, but also pose significant privacy concerns due to the potential exposure of sensitive user data, such as biometrics, behavior, identity, and real-world surroundings  \cite{S10, S15, S22, S30}. While some data points are collectible from other internet mediums (e.g., mobile phones), the unprecedented nature of the VR privacy threat stems partly from the ability to fuse a wide range of attributes that would previously have required the combined data of several devices.

\subsection{VR Attacks \& Defenses}
\label{subsec:related_work} 

There are early seminal reviews, surveys, and SoKs on mixed reality (MR) displays (1994) \cite{MR_continuum}, classifications (1996) \cite{MR_classification}, early challenges (1997) \cite{MR_challenges_I}, integrity and ownership (2000) \cite{MR_ownership_integ}, and enabler technologies (2001) \cite{MR_enablers}. 
In the 2010s, researchers delved into ethical considerations of MR (2014, 2018) \cite{MR_ethical_I, MR_Ethical_II}, updated its challenges (2016) \cite{MR_challenges_II}, discussed the threats of converging VR and social networks (2016) \cite{VR_social}, and studied VR safety (e.g., epilepsy) (2018) \cite{VR_saftey_I}.
Most recently, practitioners have continued work on VR safety attacks (e.g., misleading users to collide with their real-world surroundings) (2021, 2022) \cite{safety_II, VR_safety_manipulation}, malicious VR ads and protections \cite{N8} (2021), user authentication (2022) \cite{authentication_SoK, SOK_Auth}, and advocated for new regulations for the upcoming metaverse applications (2022) \cite{metaverse_regulation}.
While the ethics, authentication, safety, ads, regulations, and underlying technologies are paramount for VR, these works lack a focus on privacy attacks and defenses.

Our systematic search revealed the most relevant related work at the intersection of VR and privacy. 
We collected $12$ relevant literature reviews (LR) \cite{S1, S10, S12, S14, S17, S20, S24, S33, S48, S65, N3, N7}, three of which are the closest to our work.
Shrestha and Saxena (2017) \cite{S20} provided an offensive and defensive overview of eye-wears and HMDs with a focus on optical cameras in the fields of privacy, security (authentication and device integrity), and safety, with an emphasis in the latter two.
De~Guzman~et~al.~(2019) \cite{S1} expanded the augmented reality (AR) privacy and security defense classification of Roesner~et~al.~\cite{AR_protection} to MR without an in-depth analysis of data attributes and attacks.
Odeleye~et~al.~(2022) \cite{S65} provided a taxonomy of cybersecurity VR attacks related to authentication and privacy, comprising $5$ privacy defenses and $10$ attack-focused studies, which we also included in \S\ref{sec:threat_model},~\S\ref{sec:vr_attacks}, and~\S\ref{sec:vr_defenses}. 
In contrast, our work presents a more detailed exposition specific to VR and privacy, and, yet, contains a more comprehensive taxonomy of vulnerable data attributes, attacks, and defenses and a technical component on user data to highlight privacy opportunities (unlike any other LR).
Overall, extant literature examines VR and privacy as a subset of broader reviews in MR, security, and safety, thus, our detailed study focused on VR and privacy is not part of prior work.

Among the rest of selected LRs, three delved in a specific sub-field of VR and privacy, specifically, Katsini~et~al.~\cite{S14}, Kr{\"o}ger~et~al.~\cite{S33}, and Gressel~et~al.~\cite{N3} studied the privacy implications and research directions of eye-tracking.
In our work, we compiled their findings in our comprehensive taxonomies. 
Additionally, we included the relevant privacy-related insights and VR application taxonomies of two comprehensive LRs that covered general metaverse topics as varied as data management, privacy, legal issues, and economic threats \cite{S10, S24}.
Lastly, we included key information from narrower surveys in security and privacy in VR \cite{S12, S48}, and data attributes and user privacy considerations \cite{S17, N7}.

\section{Method: Data Collection \& Analysis}
\label{sec:sok_method}

The following summarizes our search approach and results described in detail in Appendix~\ref{app_sec:method}.
We employed seven of the most relevant digital libraries focused on computer science and software engineering in combination with Google Scholar to perform an exhaustive search of the extant literature.
We only included publications containing taxonomies of VR data attributes or applications, or aimed to review or implement privacy attacks or defenses in VR, from which we extracted the relevant artifacts to construct our comprehensive models and taxonomies.
With a curated set of keywords from our base literature of $12$ publications, our initial systematic search generated $1700$ hits, which we discussed and filtered, resulting in only $16$ studies. 
After deduplication, our aggregated list contained $23$ works.
We then queried their authors for additional relevant work, performed a backward search of the references of the $23$ studies, and added publications found thereafter, obtaining a final body of $74$ publications---the most recent study dates March 2023.
Further, to focus our VR attacks and defenses discussion, we designed $10$ research questions (RQ).

Lastly, we complemented the taxonomies with an analysis of VR privacy opportunities (\S\ref{sec:opportunities}) and key findings and future work items (\S\ref{sec:discussion}).
Part of the analysis quantifies some of the most sensitive and easiest-to-protect data attributes by calculating a PCA of inference attacks and weighted mean accuracy degradation after enabling the corresponding defenses.
We replicated these attacks and defenses from the most comprehensive frameworks among the $74$ studies.

\section{VR Threat \& Defense Models}
\label{sec:threat_model}

\textbf{Method}. From the $74$ selected studies, we identified $5$ studies that proposed a VR information flow \cite{S10, S15, S18, S24, S48} and $23$ with an explicit discussion or proposal for VR defense and (predominantly) threat models \cite{S1, S5, S7, S10, S15, S23, S27, S28, S29, S32, S35, S36, S37, S45, S46, S47, S50, S54, S55, S66, S67, N5, N6}.
Two researchers extracted, discussed, and combined the associated artifacts, resulting in a holistic VR information flow that frames our threat and defense models.

\begin{figure*}[htpb!]
\includegraphics[scale=0.4]{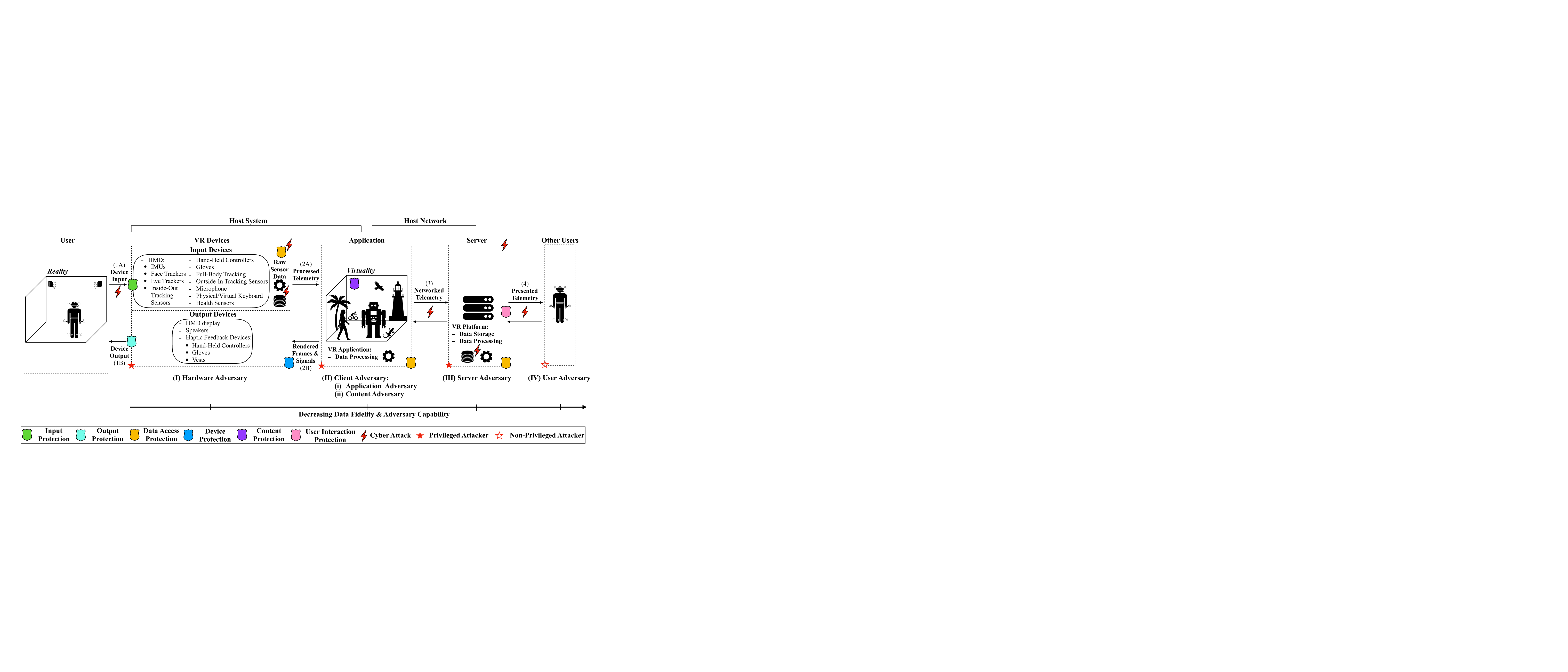}
\centering
\caption{VR information flow depicting our defense model protections, adversary model attackers, and cyber attack points. (cf. \cite{S15, S16, S18, S24}).}
\label{fig:info_flow}
\end{figure*}

\subsection{VR Information Flow}
\label{subsec:vr_info_flow}

VR device manufacturers or vendors provide app stores where users can download VR applications and games (e.g., from the Oculus Store or SteamVR).
Fig.~\ref{fig:info_flow} illustrates the information flow after installing such an application. 
These applications typically run in the host VR system, which ingests user input: \textit{geospatial \& inertial data}, \textit{audio}, \textit{text}, \textit{video}, and \textit{physiological signals} (1A).     
The various VR devices process raw sensor data and other input types into useful telemetry, which the application accesses via an API (e.g., OpenVR) (2A).
Such application controls how to use this data to generate different stimuli, e.g., visuals via a graphics rendering pipeline, audio through speakers, and haptics using hardware such as feedback vests (2B).
The output devices present this processed information to the user as an immersive, interactive virtual world (1B).
For multi-user online experiences, the client-side application exchanges processed telemetry with an external server through a network, which can reveal \textit{system} and \textit{network} user-specific information (3).
Finally, the server updates the global state of the virtual world and relays telemetry to other users (4).
As the information flows from steps (1A) to (4), intermediate data processing steps like filtering and compression degrade data quality.

\subsection{VR Threats}
\label{subsec:vr_attackers}

Within the frame of this study, we consider a state of \textit{privacy} as the lack of a breach of any individual's sensitive data attributes \cite{Wu2012}.
In our threat model, attackers breach user privacy by collecting and inferring enough information to reliably \textit{identify} and comprehensively \textit{profile} a user across VR applications over multiple usage sessions (tracking).
Attackers (i) \textit{identify} an individual when they can uniquely distinguish the user from others, and (ii) \textit{profile} users when they unwarrantedly attach information related to the user's characteristics (e.g., demographics, preferences, etc.) \cite{linddun, PriS, S52}.

The collected studies discussing or proposing threat models consider application developers \cite{S7, S15, S23, S27, S28, S29, S32, S35, S36, S37, S46, S47, S66, N5, N6}, servers \cite{S47, S15, S67, N4}, content creators \cite{S10, S15}, device manufacturers \cite{S5, S15}, other users \cite{S54, S50, S55}, and hackers\footnote{Hackers can abuse VR devices, servers, networks, or perform shoulder-surfing databases---covered extensively by security literature \cite{cyberattacks_survey_a, cyberattacks_survey_b, cyberattacks_survey_c, cyberattacks_survey_c, cyberattacks_survey_e}.} \cite{S10, S36, S45} as the attackers in VR, or rely on general privacy threat models like Lindunn \cite{linddun, PriS, S1, S23}.
Based on these studies and their system decomposition, we adopt a more comprehensive and pervasive privacy-centered attacker classification specific to VR that encompasses the privacy repercussions of the above threat models.
The adversary types of Fig.~\ref{fig:info_flow} correspond to four distinct entities associated with data processing in VR applications at different privilege levels.
These adversaries might coalesce, e.g., a developer of a VR application can also run the server providing multi-user functionality.
Table~\ref{tab:atatcker_capability} shows these attackers' capabilities.
\smallskip

\noindent \textbf{(I) Hardware Adversaries} control the hardware and firmware of the VR device and, thus, can access raw user inputs and arbitrarily manipulate the information provided to the application (2A) and presented to the user (1B).
\smallskip

\noindent \textbf{(II) Client Adversaries} represent the developers of the client-side VR application running on the VR device (\textit{Application Adversary} \cite{S15}) and the content creators (\textit{Content Adversary} \cite{S13, N8}).
Content adversaries can create \textit{immersive falsehoods}, i.e., designing immersive experiences with misinformative, manipulative, and deceptive content \cite{S13, N8}.
Application adversaries can access the input data via system APIs, and arbitrarily manipulate the rendered frames and signals output to the VR devices (2B) and the information streamed to external servers (3).
Further, the current push towards VR multitasking \cite{Meta_Multitasking}, where different apps are running concurrently, sharing the output or competing for user's inputs, increases the attack surface of the client adversary.
\smallskip

\noindent \textbf{(III) Server Adversaries} oversee the server enabling multi-user functionality and can arbitrarily process networked data before streaming it to users (4).
\smallskip

\noindent \textbf{(IV) User Adversaries} represent other users of the same VR application. They receive user data streams from a server and can interact with the target user.
\smallskip

\begin{table}[!ht]

\caption{VR attacker capabilities (cf. \cite{S15}).}
\vspace{-3mm}

\begin{center}
    
{\footnotesize

\begin{tabular}{lp{0.2mm}p{1.5mm}p{2.5mm}p{2.5mm}p{2.5mm}p{2.5mm}p{2.5mm}p{2.5mm}p{2.5mm}}

    & \multicolumn{9}{c}{\textbf{Observable Attribute Classes}} \\ \cmidrule(lr){2-10}
 
\multicolumn{1}{c}{\textbf{Adversary}} & 
\rotatebox{50}{Geo. \& Iner.} & 
\rotatebox{50}{Telemetry} &
\rotatebox{50}{Text} &
\rotatebox{50}{Audio} &
\rotatebox{50}{Video} &
\rotatebox{50}{Phy. Signals} &
\rotatebox{50}{System} &
\rotatebox{50}{Network} &
\rotatebox{50}{Behavior} 
\\ \hline

\multicolumn{1}{|l|}{\textbf{(I) Hardware}} & 
\multicolumn{2}{r}{\cmark\;\,} & \cmark & \cmark & \cmark  & \cmark & \cmark & \xmark & \multicolumn{1}{c|}{\cmark\;\,}  
\\ \hline

\multicolumn{1}{|l|}{\textbf{(II) Client}} & 
\multicolumn{2}{r}{\cmark*}  & \cmark  & \cmark & \cmark*  & \cmark & \cmark & \cmark & \multicolumn{1}{c|}{\cmark\;\,}  
\\ \hline

\multicolumn{1}{|l|}{\textbf{(III) Server}} & 
\multicolumn{2}{r}{\cmark*}  & \cmark  & \cmark* & \xmark  & \cmark & \cmark & \cmark & \multicolumn{1}{c|}{\cmark\;\,}   
\\ \hline

\multicolumn{1}{|l|}{\textbf{(IV) User}} & 
\multicolumn{2}{r}{\cmark*}  & \cmark  & \cmark* & \xmark  & \xmark  & \xmark  & \xmark  & \multicolumn{1}{c|}{\cmark\;\,}  
\\ \hline

\end{tabular}%
}
\end{center}
{\small \,\,\,\,\, *Observable in deteriorated data quality or abstracted form. \newline
\indent \,\,\,\,\, \textit{Legend}: Geo. = Geospatial, Iner. = Intertial, Phy. = Physiological. }

\label{tab:atatcker_capability}
\vspace{-5mm}
\end{table}

\subsection{VR Defenses}
\label{subsec:vr_protections}

We highlight in Fig.~\ref{fig:info_flow} where the defenses can counter potential attacks and classify them based on five adapted categories. They consist of the two categories that De~Guzman~et~al.~\cite{S1} added to the primary three proposed by Roesner~et~al.~\cite{AR_protection}, which are present in other privacy literature \cite{PETs_IoT, trask_structured_transparency_nodate, priv_review}.
Given that many researchers highlighted the potential harm of deceptive immersive content \cite{S17, S11, MR_ethical_I, MR_Ethical_II, safety_II, VR_safety_manipulation, VR_social, VR_child_abuse}, we add a category for virtual content protection.
Note that not all of these protections are related to \textit{privacy} (\S\ref{subsec:vr_attackers}), but also to \textit{security} (i.e., measures to impede unauthorized data access \cite{architecture_of_privacy}) and \textit{safety} (i.e., measures to preserve the physical and mental well-being of users \cite{S17}).
We highlight the following literature for guidance in security and safety attacks and protections: \cite{N1, N8, S1, S20, S65, S12, S24, S48, SOK_Auth, authentication_SoK}. 
We frame our SoK around attacks and defenses related to the \textit{privacy} aspects of these defenses, mainly to input protection.
\smallskip

\noindent \textbf{(I) Input Protection} (Security \& Privacy).
Software that, e.g., perturbs \cite{S16} or abstracts \cite{S37} active (user) and passive (user's surrounding environment) sensitive input information to prevent user privacy breaches.
Additionally, systems should be secured against adversarial inputs that bypass detection (cyber attack in Fig.~\ref{fig:info_flow}: 1A).
\smallskip

\noindent \textbf{(II) Data Access Protection} (Security \& Privacy). 
Active and passive user inputs are stored, relayed and accessed to deliver user-consumable output.
The corresponding privacy and security measures extensively overlap with other systems, which existing literature covers comprehensively \cite{PETs_IoT, trask_structured_transparency_nodate, priv_review, cyberattacks_survey_a, cyberattacks_survey_d}.
\smallskip

\noindent \textbf{(III) Output Protection} (Security \& Safety). 
Detecting and censoring \cite{odeleye_detecting_nodate} malicious manipulation of outputs can prevent security breaches like ``clickjacking'' \cite{AR_protection} or physical harm, e.g., inducing collisions with obscured real objects \cite{VR_safety_manipulation, safety_II}, VR sickness \cite{safety_II}, and trigger epilepsy \cite{VR_saftey_I}.
\smallskip

\noindent \textbf{(IV) User Interaction Protection} (Privacy \& Safety).
Privacy protections can enhance confidentiality (i.e., data is only revealed to selected entities \cite{PETs_IoT}) in physical or virtual spaces shared by multiple interacting users, e.g., a private virtual enclave that other users cannot enter \cite{S19}.
We add to this category safety measures such as invisible avatar barriers to avoid psychological harm from virtual harassment \cite{VR_child_abuse} or buylling \cite{S65}. 
\smallskip

\noindent \textbf{(V) Device Protection} (Security \& Safety).
Device security measures can implicitly protect users and data in the above defensive aspects, e.g., authentication prevents impersonation \cite{Luo2020OcuLockEH}, and defend against cyberattacks targeting devices \cite{S65} and networks \cite{VR_network_attack}, and VR tracking system jamming \cite{jamming_tracking}, which could lead to physical harm.
\smallskip

\noindent \textbf{(VI) Content Protection} (Privacy \& Safety).
Safety measures such as age verification and content moderation can protect users against immersive falsehoods, malicious advertisement \cite{N8}, and inappropriate, unsolicited, and harmful content that may lead to mental harm, disinformation, or manipulation of opinions and ideals \cite{VR_child_abuse, S11}.
The privacy concern involves detecting virtual content and environments nudging users to disclose sensitive information subtly, e.g., puzzles revealing health data \cite{S15}.


\section{Taxonomy of VR Data \& Applications}
\label{sec:data_taxonomy}

Thanks to the sensor-generated data and the applications processing this information, users can experience VR.
However, applications are also the gateway for adversaries to harvest sensitive user data and use such information against them. 
The following classifies and discusses the data attributes and the applications subject to our threat model.

\subsection{VR Attributes}
\label{subsec:vr_atts}

\begin{figure}[htpb!]
\includegraphics[scale=0.68]{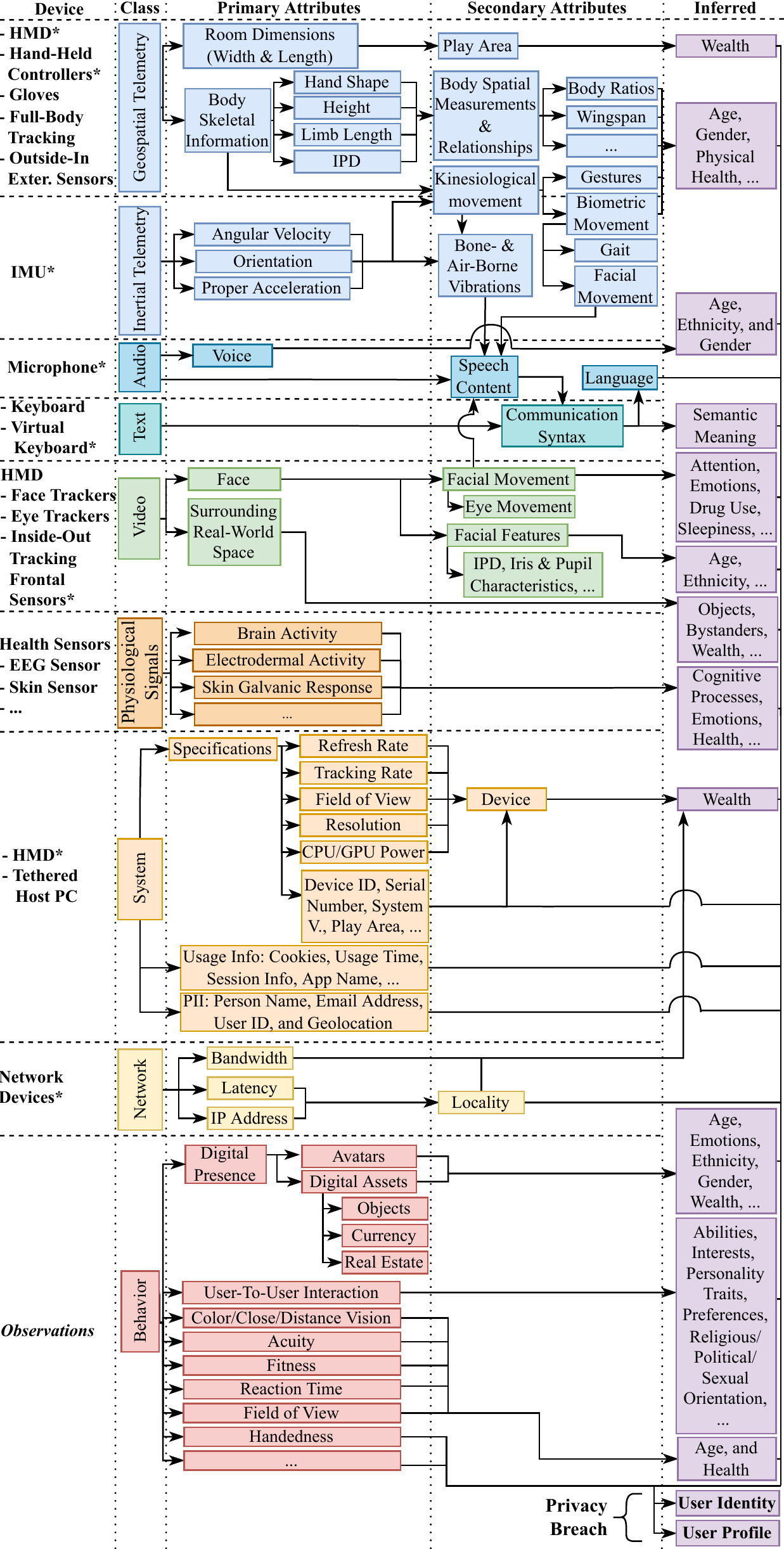}
\centering
\caption{Taxonomy of VR data attributes.\textsuperscript{*}Primary devices.}
\label{fig:data_taxonomy}
\vspace{-4mm}
\end{figure}

\textbf{Method.} We examined each of the $74$ publications to extract the highlighted, attacked, or defended data attributes that originate from users employing the input devices of \S\,\ref{subsec:vr_devices}.
Would-be attackers can collect these attributes at different steps (data sources) of the VR information flow of \S\,\ref{subsec:vr_info_flow}.
Fig.~\ref{fig:data_taxonomy} presents the resulting taxonomy of VR-derived data.
We base our categorization on observable attribute classes and indicate which attributes or observations an attacker can directly capture from a data source (\textit{primary}), deterministically derive from primary attributes (\textit{secondary}), and infer from primary and secondary attributes employing ML or other learning procedures (\textit{inferred}). 
Furthermore, we use the $74$ publications to draw the connections between attributes, thus, there might be other connections outside VR and new ones might arise in future work, e.g., deriving ethnicity or personality traits from VR inertial telemetry.
\smallskip

\noindent \textbf{Geospatial \& Inertial Telemetry.} 
The position, orientation, and acceleration of body tracking devices over time reveal anthropometric measurements.
Such measurements can be \textit{direct} (body skeletal information such as arm-length and height \cite{S30}), \textit{combined} to obtain further biometrics (e.g., wingspan \cite{S15}), or \textit{compared} to draw relationships (ratios may reveal a user's body asymmetries \cite{S16}).
An attacker may also record kinesiological movements, which can reveal unique gestures \cite{S20, S37}, or biometric movements \cite{S55} such as gait \cite{S54}.
Additionally, the devices' coordinates can map the play area's boundries, revealing its surface \cite{S15}.
Even without full-body tracking devices, Winkler~et~al.~\cite{S49} showed that reinforcement learning techniques could infer a full-body pose with telemetry from only an HMD, its IMUs, and hand-held controllers.
Furthermore, Chen~et~al.~\cite{S46} derived speech from the bone- and air-borne vibrations registered by an HMD's IMU telemetry data.
Note that hardware and client adversaries have a privileged position to observe device telemetry. 
In contrast, server and user adversaries will experience degraded precision in their attribute estimations due to intermediate data processing, e.g., filtering and compression.
\smallskip

\noindent \textbf{Audio \& Text.}
Users can verbally interact with other users in virtual telepresence applications or give voice commands to their VR devices through a microphone \cite{magic_leap}.
Attackers can listen to vocalizations to fingerprint users based on vocal characteristics (e.g., frequency or accent) \cite{S15, S20} and profile them with communication semantics \cite{S10}.
While voice biometrics may degrade along the data flow, speech semantics are more robust and could remain vulnerable to user adversaries.
Additionally, the messaging functionality enabled by physical or virtual keyboards operated with hand-held controllers or gloves increases the attack surface \cite{S66, S67, S68}. 
\smallskip

\noindent \textbf{Video.}
HMD's face optical sensors can register and track eye and facial movements and features to render expressive photorealistic avatars \cite{Codec_avatars}.
However, the facial video feed can also serve to identify an individual (e.g., using IPD, or Iris, and pupil characteristics \cite{S34, S44}) or infer emotions \cite{S3, N4}.
Notably, Kröger~et~al.~\cite{S33} provided a comprehensive overview of the plethora of attributes that privileged adversaries can infer from eye tracking. 
Moreover, with expressive avatars, server and user adversaries could also learn other users' mental state.
Additionally, while more prevalent in AR applications, the inside-out tracking frontal cameras of a VR HMD \cite{meta_quest_2} also expose the real-world environment surrounding users, which can reveal sensitive information to hardware and client adversaries, such as personal objects \cite{S23, S38}, the surrounding space type \cite{S2, S18}, or bystanders \cite{S27, S35}. 
\smallskip

\noindent \textbf{Physiological Signals.}
As health sensors like EEGs make their way into commercial-grade HMDs \cite{EEG_HMD}, the possibilities of VR (and privileged adversaries) expand dramatically. 
With these sensors, applications can adjust immersive experiences based on physiological signals that meet users' particular needs in real-time \cite{S17, S59, S61, S63} and can help users with rehabilitation treatments \cite{S20, Amelia_Virtual_Care, S60}.
Such improvements, however, will also expose critically sensitive user information, such as physical and mental health conditions \cite{S17, S63, S60}, behavior \cite{S57, S59, S62}, language semantics \cite{Meta_brain_waves, EEG_language}, and other sensitive PII like credit cards, PINs, and locations or persons known to the user \cite{BCI_Dawn}.
\smallskip

\noindent \textbf{System \& Network.}
Adversaries can determine a user's VR device, host PC, network characteristics, and related internet session information \cite{S22}. 
Specifically, hardware and client adversaries can query system APIs to collect system specifications (e.g., tracking rate, resolution, etc.), and less privileged adversaries may devise attacks to gauge a target user's refresh rate without access to system APIs or user agents \cite{S15}.
Notably, Trimananda~et~al.~\cite{S22} captured the plethora of system information relayed to servers, which included all the above, in addition to PII like a person's name and usage information such as cookies or app names.
While not specific to VR, as virtual telepresence applications rely on multiple servers to reduce perceived latency \cite{vrchat_network}, attackers can observe network traffic to determine users' geolocation without an IP address.
Altogether, these additional data points help adversaries fingerprint users to track them across internet VR sessions.
\smallskip

\noindent \textbf{Behaviour.}
Observing users' avatar likeness, expressed emotions, interactions and reactions to virtual stimuli from other avatars or virtual content can reveal various sensitive human characteristics \cite{privacy_avatars, S6, N4}. 
In practice, malicious developers may carefully and inconspicuously deliver stimuli in a virtual experience to prompt the user to unconsciously reveal their reaction time, handedness, fitness level, visual and mental acuity, etc.~\cite{S15}. 
Additionally, how a user chooses to represent their likeness as avatars, together with the digital assets they own, can reveal information such as their demographics or wealth \cite{S17, S45}.
Lastly, user-to-user interaction in social VR can lead to attackers directly spying on or engaging with the target user \cite{S19, S50}.
The information required to meaningfully observe sensitive behavioral data is typically enough at each stage of the information flow \cite{S15}.
\smallskip

\noindent \textbf{Inferred Attributes.}
With the appropriate ML algorithm \cite{evin_personality_2022, ML_twitter, ML_ontologies}, the discussed attributes above can reveal demographics \cite{S13} and other related sensitive attributes such as emotions \cite{S3, N4}, physical and mental health \cite{S33, S9}, wealth, and political or sexual orientation or preferences over different users or products \cite{S19, Jernigan_Mistree_2009}, among others \cite{S17, N7}. 
Users may also unintentionally or voluntarily self-disclose such information or additional biographical data (e.g., age, home address, education, social status, work history, etc.) \cite{S5, S6}, or be deceived by the application or other users to reveal inferable attributes \cite{S13}.
Ultimately, adversaries can leverage the breadth of data to identify and profile users across VR applications.
\smallskip

\subsection{VR Applications}
\label{subsec:vr_apps}

For decades, the gaming industry has advanced 3D graphics hardware and low-latency content delivery to create immersive, time-intensive online user experiences.
Their expertise has pushed gaming to become the current dominant application in VR \cite{the_metaverse}.
However, VR promises applications beyond entertainment: social life, education, healthcare, fitness, military training, architecture, retail, business, productivity (virtual offices), engineering, and manufacturing \cite{S21, VR_apps}.
Specifically, social VR has recently increased in popularity with titles such as \textit{VRChat}, whereby users worldwide interact with each other in real-time \cite{S6}. 

\begin{figure}[H]
\centering
\includegraphics[width=1\linewidth]{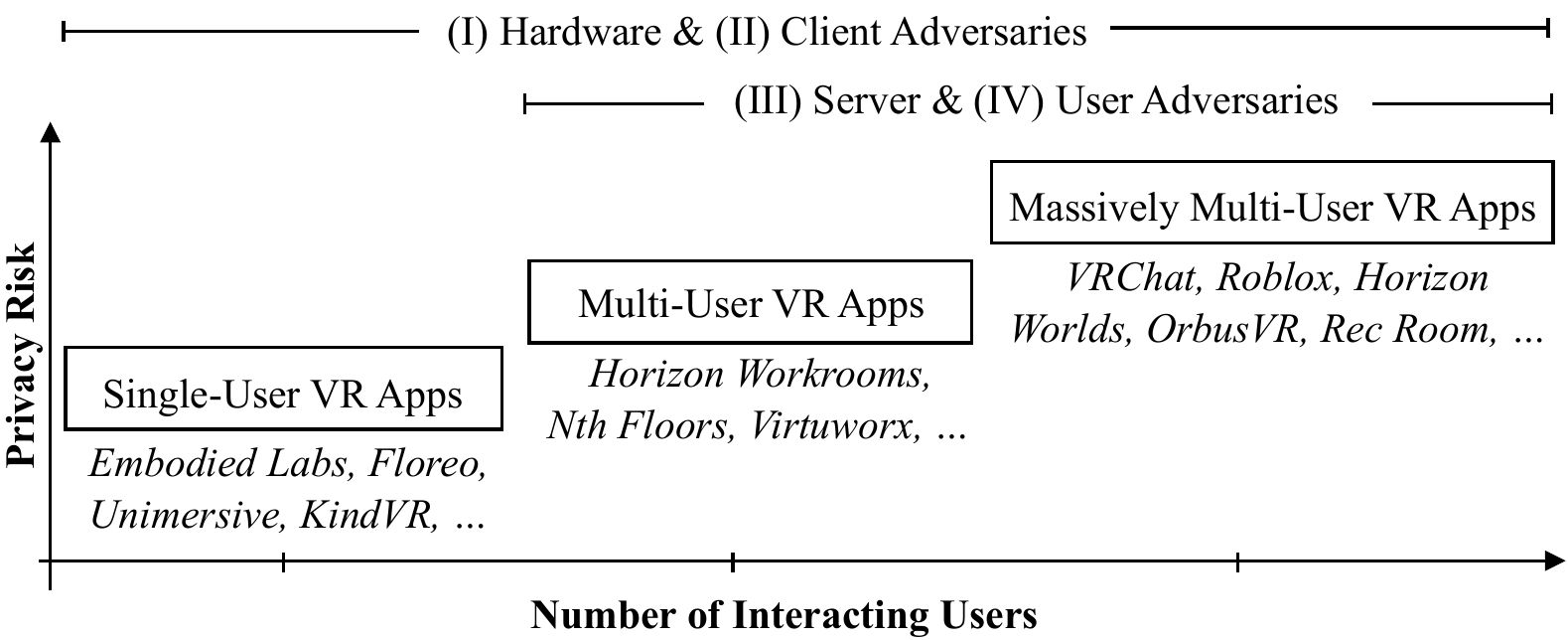}
\centering
\caption{Privacy risk of VR applications as adversary exposure increases.}
\label{fig:apps_risks}
\end{figure}

\textbf{Method.} Among the $74$ collected studies, only two classified VR applications based on the above target industries \cite{S10, S24}. 
In contrast, we provide an orthogonal categorization from a privacy standpoint inspired by \cite{S15, S1, S16} and our (i) adversary, (ii) protection models, and (iii) taxonomy of attributes.
Accordingly, we contemplate privacy risks in VR from three perspectives:
(i) \textit{adversarial}, (ii) \textit{user protections}, and (iii) \textit{data}.
VR application developers may consider answering three questions: 

(i) \textit{How much adversarial exposure could the application suffer?} 
Fig.~\ref{fig:apps_risks} shows the prevalence of hardware and client adversaries across all applications and the rise in privacy risks as users require servers to interact with others. 
While massively multi-user VR applications such as social VR are the most privacy-hostile environments, single-user applications are at least vulnerable to the VR firmware itself, as it may have direct network access to exfiltrate collected data from an application (e.g., Oculus Quest 2).
\smallskip 

(ii) \textit{How much privacy is the user willing to forgo using the application?} 
Some users are willing to expose all the information necessary to experience VR at its full immersive potential, while others are more reserved \cite{S12}.
Hence, if protecting or opting out of specific data inputs is enabled (akin to internet cookies) \cite{S16}, the privacy risks an application entails may vary from user to user. 
We suggest developers offer these protections and design their applications and games such that user experience for the privacy-conscious is not significantly deteriorated. 

\smallskip 

(iii) \textit{How sensitive is the data handled by the application?} 
Most VR applications ingest \textit{geospatial} and \textit{inertial telemetry} and \textit{audio}, and require a \textit{system} and a \textit{network} to join interactive experiences, where adversaries can extract \textit{behavioral information}. 
These attribute classes form a privacy risk baseline.
The application context raises the risks above this baseline, e.g., virtual health clinics, classrooms, and offices handle more PII and critically sensitive data than a game, e.g., \textit{physiological signals}, \textit{text} in homework or emails, and context-specific behavioural information such as attention to the lecturer or emotions during a meeting.

\section{VR Attacks}
\label{sec:vr_attacks}

\textbf{Method.} Among the $74$ collected studies, we found $34$ attacks introducing explicit, offensive mechanisms ($23$) or methods that an attacker could leverage for adversarial purposes ($11$). For example, an attacker can leverage motion device authentication software to perform identification attacks across VR sessions.
Two researchers iteratively discussed and systematically classified these attacks in Table~\ref{tab:vr_attacks} (labeled with IDs A1 to A30) based on the threat model of \S\ref{subsec:vr_attackers} and attribute classification of \S\ref{subsec:vr_atts}. 
We categorized the attacks according to the information presented in the associated papers, and included the most distinct or prevalent metrics to benchmark the attacks. 
Where information was lacking, e.g., not all attacks had an explicit adversary model, we used our best judgment supported by the publications artefacts, e.g., the client was the most common adversary and studies such as A9, A11-12 developed an application. The two researchers designed the following six RQs to focus our findings: RQ1-2 discuss opportunities for attacks, and RQ3-6 give an overview of critical attacks and explore their viability and risk.

\begin{table*}

\caption{Systematization of VR attacks from collected papers.}
\vspace{-3mm}

\begin{center}
\renewcommand{\arraystretch}{0}
\begin{tabular}{p{4mm}p{19mm}p{19mm}p{1mm}p{1mm}p{1mm}p{1mm}p{1mm}p{1mm}p{1mm}p{1mm}p{1mm}p{1mm}p{1mm}p{1mm}p{1mm}p{1mm}p{1mm}p{20mm}}
     
    &&&
    \multicolumn{9}{c}{\textbf{Observable Attribute Classes}} &
    \multicolumn{2}{c}{\textbf{PB}} &
    \multicolumn{4}{c}{\textbf{Adversary}} & 
    
    \\ \cmidrule(lr){4-18}
    
    \textbf{ID} & \textbf{Name} & \textbf{Devices} & 
    
    \rotatebox{50}{Geo.\,Tele.} & \rotatebox{50}{Inertial\,Tele.} & 
    \rotatebox{50}{Text} & \rotatebox{50}{Audio} & \rotatebox{50}{Video} & 
    \rotatebox{50}{Physio.\,Signals} & \rotatebox{50}{System} & 
    \rotatebox{50}{Network} & \rotatebox{50}{Behavior} &
    
    \rotatebox{50}{Profiling} & \rotatebox{50}{Identification} &
    
    \rotatebox{50}{Hardware} & \rotatebox{50}{Client} & 
    \rotatebox{50}{Server} & \rotatebox{50}{User} & 
    
    \textbf{Metric}
    \\ \hline 
    
    \textbf{A1} & \multicolumn{1}{|l|}{\textit{MetaData} \cite{S15}} & {\scriptsize \usym{1F576}} \faGamepad \;\faMicrophone\;\faHddO\;\faWifi &
    
    \multicolumn{1}{|l}{$\bullet$} & - & - & $\bullet$ & - & - & \multicolumn{1}{c}{$\bullet$} & $\bullet$ & \multicolumn{1}{c|}{$\bullet$} &
    
    $\bullet$ & $\bullet$ &
    
    \multicolumn{1}{|c}{-} & $\bullet$ & $\bullet$ & \multicolumn{1}{c|}{-} & 
    
    Accuracy \\ [-2.8mm]
    
    \textbf{A2} & \multicolumn{1}{|l|}{Malicious Design \cite{S13}} & N/A &
    
    \multicolumn{1}{|l}{-} & - & - & - & - & - & - & - & \multicolumn{1}{c|}{$\bullet$} & 
    
    $\bullet$ & - &
    
    \multicolumn{1}{|c}{-} & $\bullet$ & - & \multicolumn{1}{c|}{-} & 
    
    N/A \\ [-2.8mm]
    
    \textbf{A3} & \multicolumn{1}{|l|}{\textit{MitR} \cite{S50}} & N/A &
    
    \multicolumn{1}{|l}{-} & - & - & $\bullet$ & - & - & - & - & \multicolumn{1}{c|}{$\bullet$} & 
    
    $\bullet$ & $\bullet$ &
    
    \multicolumn{1}{|c}{-} & - & - & \multicolumn{1}{c|}{$\bullet$} & 
    
    N/A \\    [-2.8mm]
    
    \textbf{A4} & \multicolumn{1}{|l|}{\textit{QuestSim}\textsuperscript{$\dagger$} \cite{S49}} & {\scriptsize \usym{1F576}} \faGamepad \, \usym{1F3AF} \usym{1F518}  &
    
    \multicolumn{1}{|l}{$\bullet$} & $\bullet$ & - & - & - & - & - & - & \multicolumn{1}{c|}{-} &
    
    - & $\bullet$ &
    
    \multicolumn{1}{|c}{-} & $\bullet$ & - & \multicolumn{1}{c|}{-} & 
    
    Geo.\,Errors \\  [-2.8mm]
    
    \textbf{A5} & \multicolumn{1}{|l|}{\textit{Face-Mic} \cite{S46}} & \usym{1F3AF} \usym{1F518} &
    
    \multicolumn{1}{|l}{-} & $\bullet$ & - & - & - & - & - & - & \multicolumn{1}{c|}{-} &
    
    $\bullet$ & $\bullet$ &
    
    \multicolumn{1}{|c}{-} & $\bullet$ & - & \multicolumn{1}{c|}{-} & 
    
    Accuracy \\      [-2.8mm]
    
    \textbf{A6} & \multicolumn{1}{|l|}{ \textit{GaitLock}\textsuperscript{$\dagger$} \cite{S54}} & \usym{1F3AF} \usym{1F518} &
    
    \multicolumn{1}{|l}{-} & $\bullet$ & - & - & - & - & - & - & \multicolumn{1}{c|}{-} &
    
    - & $\bullet$ &
    
    \multicolumn{1}{|c}{$\bullet$} & - & - & \multicolumn{1}{c|}{-} & 
    
    Accuracy \\ [-2.8mm]

    \textbf{A7} & \multicolumn{1}{|l|}{Movement Biometrics\textsuperscript{$\dagger$} \cite{S11}} & {\scriptsize \usym{1F576}} \faGamepad \, \usym{1F3AF} &
    
    \multicolumn{1}{|l}{$\bullet$} & $\bullet$ & - & - & - & - & - & - & \multicolumn{1}{c|}{-} &
    
    - & $\bullet$ &
    
    \multicolumn{1}{|c}{$\bullet$} & - & - & \multicolumn{1}{c|}{-} & 
    
    EER \\   [-2.8mm]
    
    \textbf{A8} & \multicolumn{1}{|l|}{Movement Biometrics \cite{S30}} & {\scriptsize \usym{1F576}} \faGamepad \, \usym{1F3AF} &
    
    \multicolumn{1}{|l}{$\bullet$} & $\bullet$ & - & - & - & - & - & - & \multicolumn{1}{c|}{-} &
    
    - & $\bullet$ &
    
    \multicolumn{1}{|c}{-} & $\bullet$ & - & \multicolumn{1}{c|}{-} & 
    
    Accuracy \\   [-2.8mm]  
    
    \textbf{A9} & \multicolumn{1}{|l|}{Movement Biometrics\textsuperscript{$\dagger$} \cite{S51}} & {\scriptsize \usym{1F576}} \faGamepad \, \usym{1F3AF} &
    
    \multicolumn{1}{|l}{$\bullet$} & $\bullet$ & - & - & - & - & - & - & \multicolumn{1}{c|}{-}&
    
    - & $\bullet$ &
    
    \multicolumn{1}{|c}{-} & $\bullet$ & - & \multicolumn{1}{c|}{-} & 
    
    Accuracy \\    [-2.8mm]
    
    \textbf{A10} & \multicolumn{1}{|l|}{Movement Biometrics \cite{S52}} & {\scriptsize \usym{1F576}} \faGamepad \, \usym{1F3AF} \faEye  &
    
    \multicolumn{1}{|l}{$\bullet$} & $\bullet$ & - & - & $\bullet$ & - & - & - & \multicolumn{1}{c|}{-} & 
    
    $\bullet$ & $\bullet$ &
    
    \multicolumn{1}{|c}{-} & $\bullet$ & - & \multicolumn{1}{c|}{-} & 
    
    \begin{tabular}[l]{@{}l@{}} F1-Score \end{tabular}\\  [-2.8mm]
    
    \textbf{A11} & \multicolumn{1}{|l|}{Movement Biometrics\textsuperscript{$\dagger$} \cite{S53}} & {\scriptsize \usym{1F576}} \faGamepad \, \usym{1F3AF} \usym{1F518} \faEye \,  &
    
    \multicolumn{1}{|l}{$\bullet$} & $\bullet$ & - & - & $\bullet$ & - & - & - & \multicolumn{1}{c|}{-} &
    
    - & $\bullet$ &
    
    \multicolumn{1}{|c}{-} & $\bullet$ & - & \multicolumn{1}{c|}{-} & 
    
    Accuracy \\  [-2.8mm]

    \textbf{A12} & \multicolumn{1}{|l|}{ \textit{BioMove}\textsuperscript{$\dagger$} \cite{S55}} & {\scriptsize \usym{1F576}} \faGamepad \, \usym{1F3AF} \usym{1F518} \faEye &
    
    \multicolumn{1}{|l}{$\bullet$} & $\bullet$ & - & - & $\bullet$ & - & - & - & \multicolumn{1}{c|}{-} &
    
    - & $\bullet$ &
    
    \multicolumn{1}{|c}{-} & $\bullet$ & - & \multicolumn{1}{c|}{-} & 
    
    Accuracy \\  [-2.8mm]
    
    \textbf{A13} & \multicolumn{1}{|l|}{ Eye Tracking\textsuperscript{$\ddagger$} \cite{S18}} & \faEye &
    
    \multicolumn{1}{|l}{-} & -& - & - & $\bullet$ & - & - & - & \multicolumn{1}{c|}{-} &
    
    - & $\bullet$ &
    
    \multicolumn{1}{|c}{-} & $\bullet$ & - & \multicolumn{1}{c|}{-} & 
    
    \begin{tabular}[l]{@{}l@{}} Accuracy \end{tabular}\\ [-2.8mm] 
    
    \textbf{A14} & \multicolumn{1}{|l|}{ Iris Identification \textsuperscript{$\ddagger$} \cite{S34}} & \faEye &
    
    \multicolumn{1}{|l}{-} & -& - & - & $\bullet$ & - & - & - & \multicolumn{1}{c|}{-} &
    
    - & $\bullet$ &
    
    \multicolumn{1}{|c}{-} & $\bullet$ & - & \multicolumn{1}{c|}{-} & 
    
    Accuracy\\      [-2.8mm]
    
    \textbf{A15} & \multicolumn{1}{|l|}{ \textit{Kal$\varepsilon$ido}\textsuperscript{$\ddagger$} \cite{S32}} & \faEye &
    
    \multicolumn{1}{|l}{-} & -& - & - & $\bullet$ & - & - & - & \multicolumn{1}{c|}{-} &
    $\bullet$ & $\bullet$ &
    
    \multicolumn{1}{|c}{-} & $\bullet$ & - & \multicolumn{1}{c|}{-} & 
    
    \begin{tabular}[l]{@{}l@{}} F1-Score \end{tabular}\\  [-2.8mm]

    \textbf{A16} & \multicolumn{1}{|l|}{ \textit{EMOShip}\textsuperscript{$\dagger$}  \cite{S3}} & \faEye &
    
    \multicolumn{1}{|l}{-} & -& - & - & $\bullet$ & - & - & - & \multicolumn{1}{c|}{$\bullet$} &
    
    $\bullet$ & - &
    
    \multicolumn{1}{|c}{-} & $\bullet$ & - & \multicolumn{1}{c|}{-} & 
    
    \begin{tabular}[l]{@{}l@{}} F1-Score \end{tabular}\\  [-2.8mm]

    \textbf{A17} & \multicolumn{1}{|l|}{ Spatial Recognition \cite{S2}} & \faVideoCamera  &
    
    \multicolumn{1}{|l}{-} & -& - & - & $\bullet$ & - & - & - & \multicolumn{1}{c|}{-} &
    
    $\bullet$ & - &
    
    \multicolumn{1}{|c}{-} & $\bullet$ & - & \multicolumn{1}{c|}{-} & 
    
    MER \\  [-2.8mm]

    \textbf{A18} & \multicolumn{1}{|l|}{ Spatial Recognition\textsuperscript{$\ddagger$} \cite{S23}} & \faVideoCamera  &
    
    \multicolumn{1}{|l}{-} & -& - & - & $\bullet$ & - & - & - & \multicolumn{1}{c|}{-} &
    
    $\bullet$ & - &
    
    \multicolumn{1}{|c}{-} & $\bullet$ & - & \multicolumn{1}{c|}{-} & 
    
    \begin{tabular}[l]{@{}l@{}} F1-Score \end{tabular}\\  [-2.8mm]
    
    \textbf{A19} & \multicolumn{1}{|l|}{ \textit{SafeMR}\textsuperscript{$\ddagger$} \cite{S47}} & \faVideoCamera  &
    
    \multicolumn{1}{|l}{-} & -& - & - & $\bullet$ & - & - & - & \multicolumn{1}{c|}{-} &
    
    $\bullet$ & - &
    
    \multicolumn{1}{|c}{-} & $\bullet$ & $\bullet$ & \multicolumn{1}{c|}{-} & 
    
    Accuracy\\    [-2.8mm]
    
    \textbf{A20} & \multicolumn{1}{|l|}{\textit{Vreed}\textsuperscript{$\dagger$} \cite{S57}} & \faEye \;\faHeartbeat &
    
    \multicolumn{1}{|l}{-} & -& - & - & $\bullet$ & $\bullet$ & - & - & \multicolumn{1}{c|}{-} &
    
    $\bullet$ & - &
    
    \multicolumn{1}{|c}{-} & $\bullet$ & - & \multicolumn{1}{c|}{-} & 
    
    Signal Statistics\\ [-2.8mm]
    
    \textbf{A21} & \multicolumn{1}{|l|}{\textit{Galea}\textsuperscript{$\dagger$} \cite{S62}} & \faEye \;\faHeartbeat &
    
    \multicolumn{1}{|l}{-} & -& - & - & $\bullet$ & $\bullet$ & - & - & \multicolumn{1}{c|}{$\bullet$ } &
    
    $\bullet$ & - &
    
    \multicolumn{1}{|c}{$\bullet$} & - & - & \multicolumn{1}{c|}{-} & 
    
    Signal Statistics \\ [-2.8mm]
    
    \textbf{A22} & \multicolumn{1}{|l|}{Signal Processing\textsuperscript{$\dagger$} \cite{S58}} & \usym{1F3AF} \usym{1F518} \faHeartbeat &
    
    \multicolumn{1}{|l}{-} & $\bullet$ & - & - & - & $\bullet$ & - & - & \multicolumn{1}{c|}{-} &
    
    $\bullet$ & - &
    
    \multicolumn{1}{|c}{-} & $\bullet$ & - & \multicolumn{1}{c|}{-} & 
    
    EER \\ [-2.8mm]
    
    \textbf{A23} & \multicolumn{1}{|l|}{Signal Processing\textsuperscript{$\dagger$} \cite{S59}} & \usym{1F518} \faHeartbeat &
    
    \multicolumn{1}{|l}{-} & $\bullet$ & - & - & - & $\bullet$ & - & - & \multicolumn{1}{c|}{$\bullet$ } &
    
    $\bullet$ & - &
    
    \multicolumn{1}{|c}{-} & $\bullet$ & - & \multicolumn{1}{c|}{-} & 
    
    Signal Peaks\\ [-2.8mm]

    \textbf{A24} & \multicolumn{1}{|l|}{Signal Processing\textsuperscript{$\dagger$} \cite{S61}} & \faHeartbeat &
    
    \multicolumn{1}{|l}{-} & -& - & - & - & $\bullet$ & - & - & \multicolumn{1}{c|}{-} &
    
    $\bullet$ & - &
    
    \multicolumn{1}{|c}{$\bullet$} & - & - & \multicolumn{1}{c|}{-} & 
    
    Signal Stability\\ [-2.8mm]
    
    \textbf{A25} & \multicolumn{1}{|l|}{Signal Processing\textsuperscript{$\dagger$} \cite{S60}} & \faHeartbeat &
    
    \multicolumn{1}{|l}{-} & -& - & - & - & $\bullet$ & - & - & \multicolumn{1}{c|}{-} &
    
    $\bullet$ & - &
    
    \multicolumn{1}{|c}{-} & $\bullet$ & - & \multicolumn{1}{c|}{-} & 
    
    Signal Statistics\\ [-2.8mm]
    
    \textbf{A26} & \multicolumn{1}{|l|}{Signal Processing\textsuperscript{$\dagger$} \cite{S63}} & \faHeartbeat &
    
    \multicolumn{1}{|l}{-} & -& - & - & - & $\bullet$ & - & - & \multicolumn{1}{c|}{-} &
    
    $\bullet$ & - &
    
    \multicolumn{1}{|c}{-} & $\bullet$ & - & \multicolumn{1}{c|}{-} & 
    
    Accuracy \\ [-2.8mm]
    
    \textbf{A27} & \multicolumn{1}{|l|}{Signal Processing\textsuperscript{$\dagger$} \cite{S64}} & \faHeartbeat &
    
    \multicolumn{1}{|l}{-} & -& - & - & - & $\bullet$ & - & - & \multicolumn{1}{c|}{-} &
    
    $\bullet$ & - &
    
    \multicolumn{1}{|c}{-} & $\bullet$ & - & \multicolumn{1}{c|}{-} & 
    
    Signal Statistics \\ [-2.8mm]
    
    \textbf{A28} & \multicolumn{1}{|l|}{Virtual Typing \cite{S66}} & {\scriptsize \usym{1F576}} \faGamepad \, \usym{1F3AF} \usym{1F518} \scalebox{-1}[1]{\faVideoCamera}  &
    
    \multicolumn{1}{|l}{$\bullet$} & $\bullet$ & $\bullet$ & - & $\bullet$ & - & - & - & \multicolumn{1}{c|}{-} &
    
    $\bullet$ & $\bullet$ &
    
    \multicolumn{1}{|c}{$\bullet$ } & $\bullet$ & - & \multicolumn{1}{c|}{-} & 
    
    Accuracy \\ [-2.8mm]
    
    \textbf{A29} & \multicolumn{1}{|l|}{\textit{VR-Spy} \cite{S67}} & \faWifi &
    
    \multicolumn{1}{|l}{-} & - & $\bullet$ & - & - & - & - & - & \multicolumn{1}{c|}{-} &
    
    $\bullet$ & $\bullet$ &
    
    \multicolumn{1}{|c}{-} & - & $\bullet$ & \multicolumn{1}{c|}{-} & 
    
    Accuracy \\     [-2.8mm]
    
    \textbf{A30} & \multicolumn{1}{|l|}{Self-Disclosure\textsuperscript{$\dagger$} \cite{S6}} & N/A &
    
    \multicolumn{1}{|l}{-} & - & - & $\bullet$ & - & - & - & - & \multicolumn{1}{c|}{$\bullet$} &
    
    $\bullet$ & $\bullet$ &
    
    \multicolumn{1}{|c}{-} & - & - & \multicolumn{1}{c|}{$\bullet$} & 
    
    N/A \\ [-2.8mm]

    \textbf{A31} & \multicolumn{1}{|l|}{Movement Biometrics \cite{nair2023unique}} & {\scriptsize \usym{1F576}} \faGamepad  &
    
    \multicolumn{1}{|l}{$\bullet$} & - & - & - & - & - & \multicolumn{1}{c}{$\bullet$} & -& \multicolumn{1}{c|}{$\bullet$} &
    
    - & $\bullet$ &
    
    \multicolumn{1}{|c}{-} & - & $\bullet$ & \multicolumn{1}{c|}{$\bullet$} & 
    
    Accuracy \\ [-2.8mm]

    \textbf{A32} & \multicolumn{1}{|l|}{Facial Reconstruction \cite{N4}} & {\scriptsize \usym{1F576}} &
    
    \multicolumn{1}{|l}{-} & - & - & - & $\bullet$ & - & \multicolumn{1}{c}{-} & -& \multicolumn{1}{c|}{-} &
    
    $\bullet$ & - &
    
    \multicolumn{1}{|c}{-} & - & $\bullet$ & \multicolumn{1}{c|}{-} & 
    
    Accuracy \\ [-2.8mm]

    \textbf{A33} & \multicolumn{1}{|l|}{Side Channels \cite{N5}} & \faGamepad \;\faMicrophone &
    
    \multicolumn{1}{|l}{-} & - & - & - & - & - & \multicolumn{1}{c}{$\bullet$} & -& \multicolumn{1}{c|}{-} &
    
    $\bullet$ & - &
    
    \multicolumn{1}{|c}{-} & $\bullet$ & - & \multicolumn{1}{c|}{-} & 
    
    F1-Score \\ [-2.8mm]

    \textbf{A34} & \multicolumn{1}{|l|}{Movement Biometrics \cite{N6}} & {\scriptsize \usym{1F576}} \usym{1F3AF} \usym{1F518} &
    
    \multicolumn{1}{|l}{$\bullet$} & $\bullet$ & - & - & - & - & \multicolumn{1}{c}{-} & -& \multicolumn{1}{c|}{-} &
    
    $\bullet$ & - &
    
    \multicolumn{1}{|c}{-} & - & $\bullet$ & \multicolumn{1}{c|}{-} & 
    
    Top-\textit{k} Accuracy \\ [-2.8mm]
    
    \hline
    
\end{tabular}
\end{center}
{\small
\begin{flushleft}
\textsuperscript{$\dagger$}An attacker can leverage the defense/mechanism for adversarial purposes. 
\textsuperscript{$\ddagger$}Although the study is defense focused, there is an adversarial component. 
\newline 
\underline{\textit{Names}}: Names in italics correspond to the authors' selected title, otherwise, it is a descriptive name. 
\newline 
\underline{\textit{VR Device}}:  {\scriptsize \usym{1F576}} = HMD, \faEye \;\;= Eye Trackers, \faVideoCamera \;\;= Inside-Out Tracking Optical Sensors, \scalebox{-1}[1]{\faVideoCamera} = Outside-In Tracking Optical Sensors, \usym{1F3AF} = IMU Orientation, \usym{1F518} = IMU Velocity, \faGamepad \;\;= Hand-Held Controllers, \faMicrophone \; = Microphone, \faHddO \;= Tethered PC, \faWifi \;= Network Devices, \faHeartbeat \; = Health Sensor, N/A = Not Applicable.
\newline 
\hspace{-10mm} \underline{\textit{Abbreviations}}: Geo. = Geospatial, Tele. = Telemetry, Physio. = Physiological, PB = Privacy breach, MER = Mean error rate, EER = Equal error rates.
\end{flushleft}
}

\vspace{-3mm}
\label{tab:vr_attacks}
\end{table*}


\noindent \textbf{(RQ1) How can VR devices enable attack opportunities?} 
According to the literature, the most accurate identification attacks rely on \textit{HMDs} and \textit{hand-held controllers} (position and acceleration) to capture kinesiological movements (A6-12), while eye trackers mainly have a supportive role (A10-12). 
Profiling attacks that predict critically sensitive information, namely emotions (e.g., arousal and stress levels), rely on \textit{health sensors}.
These attacks use devices such as EEGs (A26), electromyograms (EMG) (A23), and ECGs (A20, A26), but also blood pressure (A25), galvanic (A20, A25-27), thermal (A24-25), respiratory (A26-27), and photoplethysmographic (A9, A22) sensors. 
Particularly, accelerometer and EMG data are an effective combination for identifying users' reactions to virtual stimuli (A23), and EEGs are especially suitable for emotion prediction (A20).
However, an attacker can also derive emotions from facial expressions reconstructed from a target ML model (A32).
Lastly, we note that some VR devices and applications have security vulnerabilities \cite{S21}, and current VR devices enable more accurate identification attacks than AR glasses (A11).

\noindent \textbf{(RQ2) How can VR data attributes expose attack opportunities?}
The \textit{geospatial telemetry} of HMDs and the hand-held controllers are a low-hanging fruit for adversaries.
Attackers can simply measure biometrics like height and wingspan to uniquely identify a small set of users (A1, $30$ users), and register unique motions such as pointing (which exhibit more identifiability than grabbing motions) and rich gestures from the dominant hand (A11).
If the application is not sand-boxed, malicious code could exploit resource monitoring and allocation APIs from the game engine to derive voice commands or hand gestures (A33).
Combined with \textit{inertial telemetry}, an attacker can infer a user's full-body pose (even with avatars of different scales, A4), inferred typed words (A34), perform highly accurate identification attacks (A7, A8, A12), and infer age (A10).
We found evidence that hand-eye coordination (A11) could provide stronger signals for identifiability than individual features, and eye-related patterns have considerable influence in gender prediction (A10), requiring the fusion with \textit{eye movement video feeds}.
In addition to proactive attacks, there is always the danger of unintentional or intentional self-disclosure (A30).
Wrt the latter, user's \textit{digital presence}, such as avatar likeness and assets, can disclose sensitive demographics, offering attack opportunities.

\noindent \textbf{(RQ3) How invasive can VR attacks be?}
The malicious accumulation of user data through profiling and tracking across internet sessions can lead to surveillance advertisement \cite{surveillance_ads, N8}, price discrimination \cite{privacyEconomicGood}, cyber abuse \cite{VR_child_abuse}, personal autonomy curtailment \cite{S17}, and pushing political agendas \cite{political_targeting}, among others \cite{MR_ethical_I, MR_Ethical_II}. These threats accentuate when adversaries can infer users' deep emotions, and reactions to stimuli \cite{S17}, which, given how immersive experiences can be (A20), are more easily observable in VR. 
Accordingly, we find critically invasive the adversarial capability to design VR experiences that can adjust users' arousal (attention) (A26) and stress (A27) to the desired level, predict anxiety (A20), and recognize emotions (A32) and their causes (A16), in the name of personalized VR experiences. 
Other attacks include generative AI producing deepfakes to create false memories (A2), and virtual content nudging users to unintentionally leak information (A1).

\noindent \textbf{(RQ4) How practical are privacy attacks?}
\textit{User} adversarial attacks are easy to execute (A3, A30), as they can, at a minimum, join a VR session and social engineer information from users. These attacks aggravate when exploitable bugs allow, e.g., invisible avatars (A3).
In contrast, attacks relying on physiological signals still require researchers to enhance their own VR \textit{hardware} to register ECGs (A22), electrodermal and muscular activity (A21), or photoplethysmographic data (A24), indicating a lack of maturity of such attacks.
Lastly, given current practices and the low establishment of VR privacy standards and enforcement \cite{S22}, an adversary in control of the \textit{client} application or the \textit{server} running multi-user functionality could easily put into practice the associated attacks the researchers have demonstrated. Most critically, they can infer in a few minutes more than $25$ attributes, including sensitive demographics (A1), hide malicious operations that collect or infer information from an otherwise honest application (A18), run emotion detection using video feeds (A16), and run authentication algorithms to tag users (A6-A12).

\noindent \textbf{(RQ5) How effective are privacy attacks?}
Based on the literature, \textit{identification} attacks targeting kinesiological movements are highly accurate.
Particularly, the most effective \textit{identification} attack targets gait and relies on dynamic time warping and sparse representation classifiers to achieve an accuracy of $98$\% using only IMUs (A6, experiments with $20$ users).
Others reach such accuracy by additionally feeding translational movements to a CNN (A7, $41$ users), in addition to eye tracking fed into kNN and SVM classifiers (A12, $15$ users).
In contrast, the most robust attack, i.e., tested with $50,000+$ participants, achieved an accuracy of $94.33$\% feeding a gradient boosting decision tree \cite{LightGBM} the positions of the HMDs and controllers of a $100$-s interval (A31).
Techniques that improve the resulting accuracy comprise normalization of height and arm lengths (A9), and smoothing methods for pre-processing (A7).

Regarding \textit{profiling}, one particularly effective and broad attack (A1, $30$ users) accurately measured from an HMD and its hand-held controllers multiple primary attributes such as height, wingspan, handedness, and interpupillary distance, among others. With this data, A1 inferred gender, age, and ethnicity with close to $100$\% accuracy, using variations of SVM and random forest classifiers.
Furthermore, the most effective emotion profiling attacks in the literature achieved an accuracy of $80$-$90$\%. They used ECG's signal amplitude and eye tracking data (A20, $12$ users) or primarily EEG (A26, $12$ users) as inputs to an SVM, or fed facial and surroundings video to tailored ML pipelines (A16, $20$ users).

\noindent \textbf{(RQ6) How these VR attacks conform with a well-established threat model?}
To conclude, we briefly examine these attacks from the perspective of a non-VR specific, highly-cited threat model: Lindunn \cite{linddun}, which guides the systematic elicitation and mitigation of privacy threats in software architectures.
All the successful attacks in Table~\ref{tab:vr_attacks} imply users' \textit{Content Unawareness} and system's \textit{Policy and Consent Non-Compliance} because the users are unaware of the hidden malicious operations that exploit the permissions granted and advertised by the VR application.
Moreover, these attacks would not be possible without incurring a \textit{Linkability} or \textit{Detectability} threat, as a successful attack must correctly link and assess the existence of a user and an attribute.
Furthermore, attacks profiling and identifying users map to the \textit{Information Disclosure} and \textit{Identifiability} threats, respectively.
Furthermore, although it was not explicitly specified, some of these attacks could also \textit{Non-Repudiate}, i.e., adversaries could show proof of the user's private virtual activities or attributes (A1, A3, A17).

\vspace{-4mm}
\section{VR Defenses}
\label{sec:vr_defenses}

\textbf{Method}. Following an equivalent method to VR attacks, we classified the $35$ identified defenses (labeled with IDs D1 to D35) according to the defense model of \S\ref{subsec:vr_protections} and attribute classification of \S\ref{subsec:vr_atts}. 
Table~\ref{tab:vr_defenses} systematically categorizes the defenses based on the corresponding papers. Similarly, we designed four RQs. While RQ1-6 draw researcher attention to areas where attacks usually excel, RQ7-10 highlight where defenses do not necessarily do: we classify defenses (RQ7) to provide a frame, and highlight usability (RQ8), limitations (RQ9), and practicality (RQ10) as focus areas for researchers---\S\ref{sec:opportunities} discusses opportunities.

\begin{table*}

\caption{Systematization of VR attacks from collected papers.}
\vspace{-3mm}

\begin{center}
\centering%
\renewcommand{\arraystretch}{0}
\begin{tabular}{p{5mm}p{20mm}p{19mm}p{1mm}p{1mm}p{1mm}p{1mm}p{1mm}p{1mm}p{1mm}p{1mm}p{1mm}p{1mm}p{1mm}p{1mm}p{1mm}p{1mm}p{1mm}p{1mm}p{1mm}p{13mm}}
     
    &&&
    \multicolumn{9}{c}{\textbf{Defended Attribute Classes}} &
    \multicolumn{2}{c}{\textbf{BP}} &
    \multicolumn{6}{c}{\textbf{Protection}} & 
    
    \\ \cmidrule(lr){4-20}
    
    \textbf{ID} & \textbf{Name} & \textbf{Devices} & 
    
    \rotatebox{50}{Geo.\,Tele.} & \rotatebox{50}{Inertial\,Tele.} & 
    \rotatebox{50}{Text} & \rotatebox{50}{Audio} & \rotatebox{50}{Video} & 
    \rotatebox{50}{Physio.\,Signals} & \rotatebox{50}{System} & 
    \rotatebox{50}{Network} & \rotatebox{50}{Behavior} &
    
    \rotatebox{50}{Profiling} & \rotatebox{50}{Identification} &
    
    \rotatebox{50}{Input} & \rotatebox{50}{Data Access} & 
    \rotatebox{50}{Output*} & \rotatebox{50}{User Interaction} &
    \rotatebox{50}{Device} & \rotatebox{50}{Content} &
    
    \textbf{Metric} 
    \\ \hline 
    
    \textbf{D1} & \multicolumn{1}{|l|}{\textit{MetaGuard} \cite{S16}} & {\scriptsize \usym{1F576}} \faGamepad \;\faMicrophone\;\faWifi &
    
    \multicolumn{1}{|l}{$\bullet$} & - & - & $\bullet$ & - & - & - & $\bullet$ & \multicolumn{1}{c|}{-} &
    
    $\bullet$ & $\bullet$ &
    
    \multicolumn{1}{|c}{$\bullet$} & - & - & - & - & \multicolumn{1}{c|}{-} &
    
    Accuracy 
    \\  [-2.8mm]
    
    \textbf{D2} & \multicolumn{1}{|l|}{Eye Tracking \cite{S18}} & \faEye &
    
    \multicolumn{1}{|l}{-} & - & - & - & $\bullet$ & - & - & - & \multicolumn{1}{c|}{-} &
    
    - & $\bullet$ &
    
    \multicolumn{1}{|c}{$\bullet$} & - & - & - & - & \multicolumn{1}{c|}{-} &
    
    F1-Score
    \\ [-2.8mm]
 
    \textbf{D3} & \multicolumn{1}{|l|}{Iris De-Identification \cite{S34}} & \faEye &
    
    \multicolumn{1}{|l}{-} & - & - & - & $\bullet$ & - & - & - & \multicolumn{1}{c|}{-} &
    
    - & $\bullet$ &
    
    \multicolumn{1}{|c}{$\bullet$} & - & - & - & - & \multicolumn{1}{c|}{-} &
    
    Accuracy
    \\ [-2.8mm]
    
    \textbf{D4} & \multicolumn{1}{|l|}{Kal$\varepsilon$ido \cite{S32}} & \faEye &
    
    \multicolumn{1}{|l}{-} & - & - & - & $\bullet$ & - & - & - & \multicolumn{1}{c|}{-} &
    
    - & $\bullet$ &
    
    \multicolumn{1}{|c}{$\bullet$} & - & - & - & - & \multicolumn{1}{c|}{-} &
    
    Accuracy
    \\ [-2.8mm]
    
    \textbf{D5} & \multicolumn{1}{|l|}{Eye Tracking \cite{S5}} & \faEye &
    
    \multicolumn{1}{|l}{-} & - & - & - & $\bullet$ & - & - & - & \multicolumn{1}{c|}{-} &
    
    $\bullet$ & $\bullet$ &
    
    \multicolumn{1}{|c}{$\bullet$} & - & - & - & - & \multicolumn{1}{c|}{-} &
    
    Accuracy
    \\ [-2.8mm]
    
    \textbf{D6} & \multicolumn{1}{|l|}{Eye Tracking \cite{S8}} & \faEye &
    
    \multicolumn{1}{|l}{-} & - & - & - & $\bullet$ & - & - & - & \multicolumn{1}{c|}{-} &
    
    $\bullet$ & $\bullet$ &
    
    \multicolumn{1}{|c}{$\bullet$} & - & - & - & - & \multicolumn{1}{c|}{-} &
    
    QoE 
    \\     [-2.8mm]
    
    \textbf{D7} & \multicolumn{1}{|l|}{Eye Tracking \cite{S7}} & \faEye &
    
    \multicolumn{1}{|l}{-} & - & - & - & $\bullet$ & - & - & - & \multicolumn{1}{c|}{-} &
    
    -& $\bullet$ &
    
    \multicolumn{1}{|c}{-} & $\bullet$ & - & - & - & \multicolumn{1}{c|}{-} &
    
    Accuracy 
    \\    [-2.8mm]     
    
    \textbf{D8} & \multicolumn{1}{|l|}{Eye Tracking \cite{S9}} & \faEye &
    
    \multicolumn{1}{|l}{-} & - & - & - & $\bullet$ & - & - & - & \multicolumn{1}{c|}{-} &
    
    $\bullet$ & - &
    
    \multicolumn{1}{|c}{-} & $\bullet$ & - & - & - & \multicolumn{1}{c|}{-} &
    
    CC, MSE
    \\    [-2.8mm] 
    
    \textbf{D9} & \multicolumn{1}{|l|}{Eye Tracking \cite{S45}} & \faEye &
    
    \multicolumn{1}{|l}{-} & - & - & - & $\bullet$ & - & - & - & \multicolumn{1}{c|}{-} &
    
    - & $\bullet$ &
    
    \multicolumn{1}{|c}{$\bullet$ } & -& - & - & - & \multicolumn{1}{c|}{-} &
    
    CRR
    \\    [-2.8mm]
    
    \textbf{D10} & \multicolumn{1}{|l|}{Eye Tracking \cite{S31}} & \faEye &
    
    \multicolumn{1}{|l}{-} & - & - & - & $\bullet$ & - & - & - & \multicolumn{1}{c|}{-} &
    
    - & $\bullet$ &
    
    \multicolumn{1}{|c}{-} & $\bullet$ & - & - & - & \multicolumn{1}{c|}{-} &
    
    CC,\;NMSE
    \\     [-2.8mm] 
    
    \textbf{D11} & \multicolumn{1}{|l|}{\textit{EyeVEIL} \cite{S44}} & \faEye &
    
    \multicolumn{1}{|l}{-} & - & - & - & $\bullet$ & - & - & - & \multicolumn{1}{c|}{-} &
    
    - & $\bullet$ &
    
    \multicolumn{1}{|c}{$\bullet$ } & -& - & - & - & \multicolumn{1}{c|}{-} &
    
    Accuracy
    \\    [-2.8mm]

    \textbf{D12} & \multicolumn{1}{|l|}{\textit{PrivacEye} \cite{S25}} & \faEye \, \faVideoCamera&
    
    \multicolumn{1}{|l}{-} & - & - & - & $\bullet$ & - & - & - & \multicolumn{1}{c|}{-} &
    
    $\bullet$ & - &
    
    \multicolumn{1}{|c}{$\bullet$ } & -& - & - & - & \multicolumn{1}{c|}{-} &
    
    Accuracy
    \\    [-2.8mm] 
 
    \textbf{D13} & \multicolumn{1}{|l|}{Spatial Recognition \cite{S23}} & \faVideoCamera&
    
    \multicolumn{1}{|l}{-} & - & - & - & $\bullet$ & - & - & - & \multicolumn{1}{c|}{-} &
    
    $\bullet$ & - &
    
    \multicolumn{1}{|c}{-} & $\bullet$ & - & - & - & \multicolumn{1}{c|}{-} &
    
    F1-Score
    \\     [-2.8mm]
    
    \textbf{D14} & \multicolumn{1}{|l|}{\textit{SafeMR} \cite{S47}} & \faVideoCamera&
    
    \multicolumn{1}{|l}{-} & - & - & - & $\bullet$ & - & - & - & \multicolumn{1}{c|}{-} &
    
    $\bullet$ & - &
    
    \multicolumn{1}{|c}{-} & $\bullet$ & - & - & - & \multicolumn{1}{c|}{-} &
    
    Accuracy
    \\    [-2.8mm]   
    
    \textbf{D15} & \multicolumn{1}{|l|}{OS Support \cite{S26, S27}} & \faVideoCamera &
    
    \multicolumn{1}{|l}{-} & - & - & - & $\bullet$ & - & - & - & \multicolumn{1}{c|}{-} &
    
    $\bullet$ & - &
    
    \multicolumn{1}{|c}{$\bullet$} & - & - & - & - & \multicolumn{1}{c|}{-} &
    
    FN, FP 
    \\     [-2.8mm]
    
    \textbf{D16} & \multicolumn{1}{|l|}{Spatial Recognition \cite{S28, S29}} & \faVideoCamera &
    
    \multicolumn{1}{|l}{-} & - & - & - & $\bullet$ & - & - & - & \multicolumn{1}{c|}{-} &
    
    $\bullet$ & - &
    
    \multicolumn{1}{|c}{$\bullet$} & - & - & - & - & \multicolumn{1}{c|}{-} &
    
    Accuracy
    \\ [-2.8mm]

    \textbf{D17} & \multicolumn{1}{|l|}{\textit{PlaceAvoider} \cite{S36}} & \faVideoCamera &
    
    \multicolumn{1}{|l}{-} & - & - & - & $\bullet$ & - & - & - & \multicolumn{1}{c|}{-} &
    
    $\bullet$ & - &
    
    \multicolumn{1}{|c}{$\bullet$} & - & - & - & - & \multicolumn{1}{c|}{-} &
    
    Accuracy
    \\  [-2.8mm]
    
    \textbf{D18} & \multicolumn{1}{|l|}{\textit{Darkly} \cite{S35}} & \faVideoCamera &
    
    \multicolumn{1}{|l}{-} & - & - & - & $\bullet$ & - & - & - & \multicolumn{1}{c|}{-} &
    
    $\bullet$ & $\bullet$ &
    
    \multicolumn{1}{|c}{$\bullet$} & - & - & - & - & \multicolumn{1}{c|}{-} &
    
    \#\;Breaches
    \\  [-2.8mm]
    
    \textbf{D19} & \multicolumn{1}{|l|}{Spatial Recognition \cite{S38}} & \faVideoCamera &
    
    \multicolumn{1}{|l}{-} & - & - & - & $\bullet$ & - & - & - & \multicolumn{1}{c|}{-} &
    
    $\bullet$ & - &
    
    \multicolumn{1}{|c}{$\bullet$} & - & - & - & - & \multicolumn{1}{c|}{-} &
    
    Accuracy
    \\  [-2.8mm]
    
    \textbf{D20} & \multicolumn{1}{|l|}{Spatial Recognition \cite{S39}} & \faVideoCamera &
    
    \multicolumn{1}{|l}{-} & - & - & - & $\bullet$ & - & - & - & \multicolumn{1}{c|}{-} &
    
    - & $\bullet$ &
    
    \multicolumn{1}{|c}{$\bullet$} & - & - & - & - & \multicolumn{1}{c|}{-} &
    
    N/A
    \\  [-2.8mm]
    
    \textbf{D21} & \multicolumn{1}{|l|}{\textit{OpenFace} \cite{S40}} & \faVideoCamera &
    
    \multicolumn{1}{|l}{-} & - & - & - & $\bullet$ & - & - & - & \multicolumn{1}{c|}{-} &
    
    - & $\bullet$ &
    
    \multicolumn{1}{|c}{$\bullet$} & - & - & - & - & \multicolumn{1}{c|}{-} &
    
    Accuracy
    \\  [-2.8mm]
    
    \textbf{D22} & \multicolumn{1}{|l|}{\textit{GARP-Face} \cite{S41}} & \faVideoCamera &
    
    \multicolumn{1}{|l}{-} & - & - & - & $\bullet$ & - & - & - & \multicolumn{1}{c|}{-} &
    
    - & $\bullet$ &
    
    \multicolumn{1}{|c}{$\bullet$} & - & - & - & - & \multicolumn{1}{c|}{-} &
    
    Accuracy
    \\  [-2.8mm]
    
    \textbf{D23} & \multicolumn{1}{|l|}{GAN-Based Defense \cite{S42}} & \faVideoCamera &
    
    \multicolumn{1}{|l}{-} & - & - & - & $\bullet$ & - & - & - & \multicolumn{1}{c|}{-} &
    
    - & $\bullet$ &
    
    \multicolumn{1}{|c}{$\bullet$} & - & - & - & - & \multicolumn{1}{c|}{-} &
    
    Natruralness
    \\     [-2.8mm] 
    
    \textbf{D24} & \multicolumn{1}{|l|}{GAN-Based Defense \cite{S43}} & \faVideoCamera &
    
    \multicolumn{1}{|l}{-} & - & - & - & $\bullet$ & - & - & - & \multicolumn{1}{c|}{-} &
    
    - & $\bullet$ &
    
    \multicolumn{1}{|c}{$\bullet$} & - & - & - & - & \multicolumn{1}{c|}{-} &
    
    Accuracy
    \\     [-2.8mm]
    
    \textbf{D25} & \multicolumn{1}{|l|}{\textit{Prepose} \cite{S37}} & \scalebox{-1}[1]{\faVideoCamera} &
    
    \multicolumn{1}{|l}{$\bullet$} & - & - & - & $\bullet$ & - & - & - & \multicolumn{1}{c|}{-} &
    
    - & $\bullet$ &
    
    \multicolumn{1}{|c}{$\bullet$} & - & - & - & - & \multicolumn{1}{c|}{-} &
    
    Expression
    \\  [-2.8mm]
    
    \textbf{D26} & \multicolumn{1}{|l|}{Movement Biometrics \cite{S51}} & {\scriptsize \usym{1F576}} \faGamepad \; \usym{1F3AF}  &
    
    \multicolumn{1}{|l}{$\bullet$} & $\bullet$ & - & - & - & - & - & - & \multicolumn{1}{c|}{-} &
    
    - & $\dagger$ &
    
    \multicolumn{1}{|c}{-} & - & - & - & $\bullet$ & \multicolumn{1}{c|}{-} &
    
    Accuracy
    \\  [-2.8mm]
    
    \textbf{D27} & \multicolumn{1}{|l|}{Movement Biometrics \cite{S53}} & {\scriptsize \usym{1F576}} \faGamepad \; \usym{1F3AF} \usym{1F518} \faEye &
    
    \multicolumn{1}{|l}{$\bullet$} & $\bullet$ & - & - & $\bullet$ & - & - & - & \multicolumn{1}{c|}{-} &
    
    - & $\dagger$ &
    
    \multicolumn{1}{|c}{-} & - & - & - & $\bullet$ & \multicolumn{1}{c|}{-} &
    
    Accuracy
    \\  [-2.8mm]

    \textbf{D28} & \multicolumn{1}{|l|}{\textit{GaitLock} \cite{S54}} & \usym{1F3AF} \usym{1F518} &
    
    \multicolumn{1}{|l}{-} & $\bullet$ & - & - & - & - & - & - & \multicolumn{1}{c|}{-} &
    
    - & $\dagger$ &
    
    \multicolumn{1}{|c}{-} & - & - & - & $\bullet$ & \multicolumn{1}{c|}{-} &
    
    Accuracy
    \\   [-2.8mm]
    
     \textbf{D29} & \multicolumn{1}{|l|}{\textit{BioMove} \cite{S55}} & {\scriptsize \usym{1F576}} \faGamepad \; \usym{1F3AF} \usym{1F518} \faEye &
    
    \multicolumn{1}{|l}{$\bullet$ } & $\bullet$ & - & - & $\bullet$ & - & - & - & \multicolumn{1}{c|}{-} &
    
    - & $\dagger$ &
    
    \multicolumn{1}{|c}{-} & - & - & - & $\bullet$ & \multicolumn{1}{c|}{-} &
    
    Accuracy
    \\      [-2.8mm] 
    
    \textbf{D30} & \multicolumn{1}{|l|}{Digital Presence \cite{S19}} & N/A &
    
    \multicolumn{1}{|l}{-} & - & - & - & - & - & - & - & \multicolumn{1}{c|}{$\bullet$} &
    
    $\bullet$ & $\bullet$ &
    
    \multicolumn{1}{|c}{-} & - & - & $\bullet$ & - & \multicolumn{1}{c|}{$\bullet$} &
    
    N/A
    \\    [-2.8mm]  

    \textbf{D31} & \multicolumn{1}{|l|}{\textit{SecSpace} \cite{S56}} & N/A &
    
    \multicolumn{1}{|l}{-} & - & - & - & - & - & - & - & \multicolumn{1}{c|}{$\bullet$} &
    
    - & $\bullet$ &
    
    \multicolumn{1}{|c}{-} & - & - & $\bullet$ & - & \multicolumn{1}{c|}{-} &
    
    N/A
    \\     [-2.8mm] 
    
    \textbf{D32} & \multicolumn{1}{|l|}{Design Defense \cite{S13}\textsuperscript{$\ddagger$}} & N/A &
    
    \multicolumn{1}{|l}{-} & - & - & - & - & - & - & - & \multicolumn{1}{c|}{$\bullet$} &
    
    $\bullet$ & - &
    
    \multicolumn{1}{|c}{-} & - & - & $\bullet$ & - & \multicolumn{1}{c|}{$\bullet$} &
    
    N/A
    \\ [-2.8mm]
    
    \textbf{D33} & \multicolumn{1}{|l|}{\textit{MitR Defense} \cite{S50}\textsuperscript{$\ddagger$}} & N/A &
    
    \multicolumn{1}{|l}{-} & - & - & $\bullet$ & - & - & - & - & \multicolumn{1}{c|}{$\bullet$} &
    
    $\bullet$ & $\bullet$ &
    
    \multicolumn{1}{|c}{-} & - & - & $\bullet$ & - & \multicolumn{1}{c|}{-} &
    
    FN, FP
    \\    [-2.8mm] 

    \textbf{D34} & \multicolumn{1}{|l|}{Self-Disclosure Defense \cite{S4}} & N/A &
    
    \multicolumn{1}{|l}{-} & - & -& $\bullet$ & - & - & - & - & \multicolumn{1}{c|}{$\bullet$} &
    
    $\bullet$ & $\bullet$ &
    
    \multicolumn{1}{|c}{-} & - & - & $\bullet$ & - & \multicolumn{1}{c|}{-} &
    
    N/A
    \\   [-2.8mm]   
    
    \textbf{D35} & \multicolumn{1}{|l|}{\textit{ReconViguRation} \cite{S68}} & \faKeyboardO &
    
    \multicolumn{1}{|l}{-} & - & $\bullet$ & - & - & - & - & - & \multicolumn{1}{c|}{-} &
    
    $\bullet$ & $\bullet$ &
    
    \multicolumn{1}{|c}{$\bullet$} & - & - & - & - & \multicolumn{1}{c|}{-} &
    
    Error Rate
    \\    [-2.8mm] 
    
    \hline
    
\end{tabular}
\end{center}

{\small
\begin{flushleft}
\textsuperscript{$\ddagger$}Although the study is attack focused, there is a defensive component. \textsuperscript{$\dagger$}Authentication protection (as opposed to identification protection). 
 
*Output safety and security attacks and defenses are covered in dedicated literature \cite{AR_protection, VR_safety_manipulation, safety_II, odeleye_detecting_nodate, VR_saftey_I}. 

\underline{\textit{Names}}: Names in italics correspond to the authors' selected title, otherwise, it is a descriptive name. 

\underline{\textit{VR Device}}:  {\scriptsize \usym{1F576}} = HMD, \faEye \;\;= Eye Trackers, \faVideoCamera \;\;= Inside-Out Tracking Optical Sensors, \scalebox{-1}[1]{\faVideoCamera} = Outside-In Tracking Optical Sensors, \usym{1F3AF} = IMU Orientation, \usym{1F518} = IMU Velocity, \faGamepad \;\;= Hand-Held Controllers, \faMicrophone \; = Microphone, \faWifi \;= Network Devices, \faKeyboardO \;= Physical keyboard; N/A = Not Applicable/Available.

\underline{\textit{Abbreviations}}: Geo. = Geospatial, Tele. = Telemetry, Physio. = Physiological, BP = Breach prevention, QoE = Quality of experience, CC = Correlation coefficient, (N)MSE = (Normalized) Mean square error, CRR = Correct recognition rate, FN = False negative, FP = False positive. 
\end{flushleft}
}

\label{tab:vr_defenses}
\vspace{-3mm}
\end{table*}


\noindent \textbf{(RQ7) What are the types of defensive mechanisms?}

(i) \textit{Perturbation}. 
Some works provide provable privacy guarantees by adding noise (differential privacy (DP) \cite{DP-original}) to geospatial or eye tracking data (e.g., D1, D4), while others blur (e.g., D11, D18) or mask (e.g., D3, D17) regions of a video like facial features, sensitive objects, and bystanders.

(ii) \textit{Information abstraction}. 
Software that extracts key features from the surrounding space (e.g., surfaces, D16) or shares only the events triggered by sensitive inputs (e.g., unique gestures, D25). 

(iii) \textit{Recognizers}. 
Automated deepfake detection (D32), and middleware that detects and warns the user of sensitive surrounding objects and bystanders (D15). 

(iv) \textit{Static \& Dynamic Analyzers}.  
Detection of application vulnerabilities that could lead to, e.g., unauthorized access to a private VR room (D33), and malware that, e.g., detects and exfiltrates sensitive surrounding objects (D13).

(v) \textit{Platform features}. 
These solutions comprise mainly user-interaction protections. The primary examples include virtual (and physical) private enclaves that only authorized users can trespass (D31). Moreover, other defenses focus on confusing adversaries, e.g., with avatar clones dispersed across multiple VR applications, teleportation to new virtual locations, private copies of the virtual public environment, and platform-generated non-identifiable or invisible avatars (D30). Furthermore, platforms could include embedded voice modulators and social media privacy settings, whereby, e.g., only friends could see one's avatar (D32). 

(vi) \textit{Authentication}. 
Biometric movement recognition for logging into a VR device (D26-29).

\noindent \textbf{(RQ8) How defenses balance usability and privacy?}
Usability is critical for immersion in VR applications; thus, researchers design utility metrics to assess the loss of usability when the user enables privacy protections.
Aspects that impact usability are battery \textit{energy consumption} (D14, D17), \textit{latency} (D2, D4), and \textit{playability} (D1, D4, D9, D14), i.e., how enjoyable or productive a VR experience is.
Approaches that help to minimize device \textit{energy consumption} are a tethered PC, offloading computation to the cloud (although bandwidth may become a challenge, D17), and sharing processing resources like object detection with other applications, which also reduces \textit{latency} (D14).
VR protections can decrease \textit{playability} if the defense perturbs data, which is measurable primarily with metrics such as game scores (D1), subjective enjoyment (D4), attentiveness, comfort (D9), naturalness (D23), and accuracy loss (D1, D4, D16), among others.
Additionally, enabling users to manage their privacy makes protections more usable (D31).
Hence, applications empower users by giving them a choice (and the responsibility) to switch protections \textit{on} depending on the context (D11), select their privacy strength with modulators like sliders (D1, D4), and providing visual prompts that communicate the impact of applications accessing user data (D15). In the background, these choices change the parameters quantifying privacy (e.g., $\varepsilon$ in DP), values which we suggest setting empirically ex-ante (D1-2, D4, D5). 

\noindent \textbf{(RQ9) What are the limitations in privacy defenses?}
Based on the literature, we suggest several key improvement areas: (i) Update \textit{perturbation} protections that use generative adversarial networks, considering the rapid advancements in generative AI \cite{Drobyshev22MP}, as the latest study available is almost four years old (D24). (ii) Enhance biometric-movement authentication to cover activities beyond walking, like running (D28). (iii) Expand provable privacy guarantees to eye tracking (D4, D8, D10) and geospatial (D1) time series. (iv) Increase research on user privacy preferences in VR to contextualize protections, (v) improve permission structures of VR operating systems (D13), and (vi) develop a standard vetting program to verify VR library functions (D16). However, the closed-source nature of VR systems \cite{meta_quest_pro} may limit defensive deployments.


\noindent \textbf{(RQ10) How practical and effective are privacy defenses?}
Researchers typically implement defenses as middleware (D15-16, D18) that pre-processes data before a potentially malicious application ingests it, or as an easy-to-install plugin within the application (D1). In terms of effectiveness, the latter would only defend against \textit{server} and \textit{user} adversaries. Further, unless the implementation is at the firmware level, users have no protection against a \textit{hardware} adversary.
Additionally, we identified in the literature promising prototypical defenses that significantly reduced the accuracy of \textit{identification} attacks based primarily on geospatial (D1) and eye tracking (D4) data to random-guess levels. Moreover, we found works demonstrating substantial accuracy degradation in \textit{profiling} attacks at reasonable privacy levels (DP with $3 \leq \varepsilon \leq 6$). Notable examples include \textit{gender} inference based on eye tracking, whose accuracy D5 reduced to random guessing, and deriving \textit{age} primarily from geospatial telemetry, which D1 reduced by $58.25$\%.

\section{VR Privacy Opportunities}
\label{sec:opportunities}

This section complements the taxonomies by exploring privacy practices and opportunities based on a quantitative analysis of attacks and defenses (Tables~\ref{tab:pca} and~\ref{tab:MetaGuard_impact}), and a qualitative examination of research gaps (Table~\ref{tab:min_set}), and in practice.

\subsection{Quantitative Analysis}
\label{subsec:quantitative_analysis}

Our primary objective is not to verify the results of the studies, instead, it illustrates a possible method of ranking attributes based on their protectability, with an intent to inspire researchers to routinely incorporate this analysis in their investigations. 
To accomplish this, we searched for pairs of complementary open-source works (one attack and one defense) that considered a wide range of granular attributes that could be ranked in terms of privacy risk and defensibility.
Within the limited pool of only four open-source defense-focused works, only \textit{MetaGuard} (D1) \cite{metaguard_repo} satisfied our criteria, primarily due to its unambiguous coupling with an attack study ((A1) \textit{MetaData} \cite{game_repo}).
While we have corroborated these studies' results, note that they still await peer-review.

\begin{table*}[htpb!]

\caption{Cumulative explained variability (\%) of PC1 for each primary and secondary attribute per inferred attribute.}

\vspace{-4mm}
\begin{center}
{\scriptsize
\begin{tabular}{|l||p{9mm}|c|p{7mm}|c|c|p{8mm}|c|c|c|c|c|c|}
\hline 
\multicolumn{1}{|c||}{\textbf{Inferred}}  & \multicolumn{12}{c|}{\textbf{Primary} \& \textbf{Secondary Attributes}} 
\\ \hline 

\textbf{Demographics} & \centering  \textbf{Height} & \thead{\textbf{Room} \\ \textbf{Size}} & \centering \textbf{Voice} &  \textbf{Wingspan} & \thead{\textbf{Longer} \\ \textbf{Arm}} & \centering  \textbf{IPD} & \textbf{Vision} & \thead{\textbf{Reaction} \\ \textbf{Time}} & \thead{\textbf{Game} \\ \textbf{Duration}} & \textbf{Location} & \thead{\textbf{MoCA} \\ \textbf{Score}} &  \textbf{Language} \\ \hline\hline
  
\textbf{Gender} & \centering $51.07$\% & -- & \centering -- & $33.65$\% & -- & \centering  $15.28$\% & -- & -- & -- &  -- & -- & --
\\ \hline

\textbf{Age} & \centering $26.63$\% & -- & -- & \centering -- & -- & \centering  -- & $25.66$\%  & $21.44$\%  & $17.03$\% & --  &  $9.24$\% & --
\\ \hline

\textbf{Ethnicity} & \centering $14.19$\% & -- & \centering $84.68$\% & -- & -- & \centering  -- & -- & -- & -- &  -- & -- & $1.13$\%
\\ \hline

\textbf{Income} & \centering -- & $85.76$\% & \centering -- & -- & -- & \centering  -- & -- & -- & -- & $14.24$\% &  -- & --
\\ \hline

\textbf{Identity} & \centering $21.29$\% & -- & \centering -- & $21.37$\% & $37.08$\% & \centering  $20.25$\% & -- & -- & -- & -- &  -- & --
\\ \hline \hline

\textbf{Risk} \hfill \textit{Total}: & \centering $113.18$\% & $85.76$\% & \centering $84.68$\% & $55.03$\% & $37.08$\% & \centering $35.53$\% & $25.66$\% & $21.44$\% & $17.03$\% &  $14.24$\% & $9.24$\% & $1.13$\%
\\ \hline

\end{tabular}%
}
\end{center}

\label{tab:pca}
\vspace{-3mm}
\end{table*}

\noindent \textbf{Evaluation Method.}
Tables~\ref{tab:pca} and~\ref{tab:MetaGuard_impact} present the results of this analysis, which consists of three steps: 

(i) \textit{Risk.}
We ran a PCA with Azure ML~\cite{azure_ml} over the anonymized ground truth of the participants of the \textit{MetaData} study to calculate the amount of variability explained in PC1 by each attribute (e.g., height, or wingspan) for each inferred data point (e.g., gender, or age).
Summing attribute contributions yields a summary statistic of the risk of attribute leakage, representing the information adversaries could obtain from their observations (Table~\ref{tab:pca}).

(ii) \textit{Weighted Mean Degradation}. 
For this metric, we relied on the anonymized frame-by-frame telemetry data of the $30$ participants of the \textit{MetaData} study, which we used to replicate the attacks with nearly identical accuracy.
The replicated attacks measured the sensitive target attributes.
Consecutively, we repeated the attacks with \textit{MetaGuard} enabled at three privacy levels: low, medium, and high, to measure the degraded new attack accuracy.
With these results, we performed a weighted average on the degraded attack accuracy at different privacy protection levels to reveal the attributes easiest to protect, i.e., with the highest accuracy degradation (Table~\ref{tab:MetaGuard_impact}).  

(iii) \textit{Opportunity.}
Finally, we ordered the attributes with the highest accuracy degradation and risk in Table~\ref{tab:MetaGuard_impact}, highlighting the most sensitive and easiest-to-protect attributes.

\begin{table*}[htpb!]

\caption{Weighted mean degradation of attack accuracy per attribute using \textit{MetaGuard} \cite{S16} as defense mechanism.}
\vspace{-4mm}

\begin{center}
{\scriptsize
\begin{tabular}{|l|c||cc|cc|cc||c|c||c|}
\hline
\textbf{Attribute}  & \thead{\textbf{Accu. at} \\ \textbf{No Priv.}} &
\thead{\textbf{L Priv.} \\ \textbf{Parameter}} & 
\thead{Accu. at \\ \textbf{L Priv.}} &  
\thead{\textbf{M Priv.} \\ \textbf{Parameter}} & 
\thead{Accu. at \\ \textbf{M Priv.}} &  
\thead{\textbf{H Priv.} \\ \textbf{Parameter}}  & 
\thead{Accu. at \\ \textbf{H Priv.}} &  
\textbf{WMD} & 
\textbf{Risk} & 
\thead{ \textbf{Opportunity} \\ \textbf{(Score)}}
\\ \hline \hline

\textbf{Room Size}	& 97\%	&	$\varepsilon$=3.00 &	33.52\% &	$\varepsilon$=1.00	&	23.44\%	&	$\varepsilon$=0.10	&	19.53\%		&	66.28\%  & 85.76\% & \textbf{56.84}
\\ \hline

\textbf{Height} & 100\%	&	$\varepsilon$=5.00 & 68.6\%	&	$\varepsilon$=3.00 &	58.17\%	&	$\varepsilon$=1.00	&	44.47\%		&	37.56\% & 113.18\% & \textbf{42.51}
\\ \hline

\textbf{IPD} & 87\% &	$\varepsilon$=5.00 &	19.47\%	&	$\varepsilon$=3.00 & 14.17\%	&	$\varepsilon$=1.00	&	12.17\%		&	
70.11\% & 35.53\% & \textbf{24.91}
\\ \hline

\textbf{Wingspan} & 100\%		&	$\varepsilon$=3.00 &	78.80\%		&	$\varepsilon$=1.00  &	66.00\%	&	$\varepsilon$=0.50 & 	65.46\%		&	25.53\% & 55.03\%  & \textbf{14.05}
\\ \hline

\textbf{Location} & 90\%	&	$d$=25.00 ms &	6.66\% &	$d$=30.00 ms	&	0.00\%	&	$d$=50.00 ms	&	0.00\%		&	88.41\% & 14.24\% & \textbf{12.59}
\\ \hline

\textbf{Voice}	& 63\%	&	$\varepsilon$=6.00 &	52.50\%	 &	$\varepsilon$=1.00 &	40.00\%	&	$\varepsilon$=0.10	& 32.50\%		&	12.54\% & 84.68\% & \textbf{10.62}
\\ \hline

\textbf{Longer Arm}	& 100\%	&	$\varepsilon$=3.00 &	77.78\%	&	$\varepsilon$=1.00 &	62.22\%	&	$\varepsilon$=0.10	&	53.33\%		&	26.61\% & 37.08\% & \textbf{9.87}
\\ \hline

\textbf{Reaction Time} & 87.50\%	&	$d$=10.00 ms &	79.20\%		&	$d$=20.00 ms	&	62.50\%	&	$d$=100.00 ms &	54.20\%		&	30.10\%  & 21.44\% & \textbf{6.45}
\\ \hline

\textbf{Refresh Rate} & 100\%		&	$d$=90.00 ms &	0.00\%	&	$d$=72.00 ms	&	0.00\%	&	$d$=60.00 ms &	0.00\%		&	100.00\% & -- & --
\\ \hline

\textbf{Tracking Rate} & 100\%		&	$d$=90.00 ms &	0.00\% &	$d$=72.00 ms	&	0.00\%	&	$d$=60.00	ms	&	0.00\%	 &	100.00\% & -- & --
\\ \hline

\textbf{Handedness}	 & 97\%	&	$\varepsilon$=1.30 &	92.50\%	&	$\varepsilon$=1.00 &	75.00\%	&	$\varepsilon$=0.70	&	57.50\%		&	18.50\% & -- & --
\\ \hline

\textbf{Physical Fitness} & 90.00\%		&	$\varepsilon$=5.00 &	86.11\%	&	$\varepsilon$=3.00 &	79.11\%	&	$\varepsilon$=1.00	&	61.56\%		&	8.95\% & -- & --
\\ \hline

\end{tabular}
}
\end{center}

{\small
\hspace{-20mm} \underline{\textit{Abbreviations}}: $d$ = delay, Accu. = Accuracy, L = Low, M = Medium, H = High, Priv. = Privacy, \newline  
\;\;\; WMD = Weighted Mean Degradation = $(Accuracy$ $at$ $No$ $Privacy)$ - $ \tfrac{\sum_{i \in \{L, M, H\}} \left(Paramater_i*Accuracy_i\right)}{\sum_{i \in \{L, M, H\}} \left(Paramater_i\right)} $, Opportunity = WMD $*$ Risk.
}

\label{tab:MetaGuard_impact}
\vspace{-4mm}
\end{table*}

\noindent \textbf{Opportunities.}
We suggest privacy practitioners to prioritize deploying differential privacy defenses that protect \textit{room size}, \textit{height}, and \textit{interpupillary distance} (IPD) in their devices and applications, as they show the highest leakage risk and sensitivity to noise.
\vspace{0.25mm}

\noindent \textbf{Ethical Considerations}
This SoK does not contain original data collected from human subjects. 
We replicated prior studies employing anonymous data collected directly from the authors of those studies or publicly available online repositories.
We verified that OHRP-registered institutional ethics review boards processed and approved those prior studies and considered them non-deceptive.
Additionally, those studies' informed consent documents included permission to re-use collected data for follow-up research work, and we handled such data rigorously according to the corresponding original consent documentation requirements.

\subsection{Min-Set Coverage of Defenses \& Attacks}
\label{subsec:min_set}

\textbf{Method.} We selected the most comprehensive attacks per attribute class and mapped them to the most fitting defenses in Table~\ref{tab:min_set}, considering identification and profiling privacy breaches and highlighting privacy opportunities. 

On the one hand, while most of the attacks had an associated defense, some protections did not defend the full spectrum of the attack vector.
For instance, audio identification protection has seen more defenses outside of VR \cite{nautsch_preserving_2019}, (D35) ReconViguRation did not account for (A29) \textit{VR-Spy} (which leverages the channel state information of WiFi signals to infer unique gesture patterns), and (D1) \textit{MetaGuard} could not protect against malicious content, with which (A1) \textit{MetaData} covertly assessed the cognitive abilities.
Thus, we encourage researchers to re-examine the proposed defenses and improve them for completeness.
On the other hand, we highlight in Table~\ref{tab:min_set} larger research gaps in the landscape of VR threats and protections as opportunities for researchers to advance the field.
Primarily, despite existing orthogonal research \cite{Meta_brain_waves, EEG_language, BCI_Dawn}, physiological signals are understudied from an adversarial and defense perspective, specifically in the context of VR.
There is a dearth of defenses against using geospatial and inertial data for identification and profiling (e.g., A5-6, A10, A12), and research on usability impact of defenses, like blurring bystanders effect on user recollection \cite{N2}.

\subsection{VR Defenses in Practice}
\label{subsec:priv_sens_apps}

\textbf{Method.} We examined the $58$ studies on attacks and defenses to find open-source implementations of their proposed mechanisms, and briefly explored privacy threats and protections in the industry.

Unfortunately, we found that only $21$\% had a \textit{functional} repository: $4$ defenses (D1, D16, D25, D33) and $8$ attacks (A1, A7-8, A16-17, A21, A27, A31).
Despite their defensive efforts, e.g., (D1) \textit{MetaGuard} packaged the first VR ``incognito mode'' as a Unity plugin \cite{metaguard_repo} for any VR application using MelonLoader \cite{melon_loader}, only (D25) \textit{Prepose} had an official affiliation with Microsoft (with no evidence of its use in production), and the Bigscreen company used the recommendations from (D33) \textit{MitR Defense} to patch their privacy vulnerabilities.
On the other hand, (A16) \textit{EMOShip} forms part of the technology stack of Pupil Labs.
Overall, we observe little transfer of privacy research into the VR industry.
Additionally, the conclusions from the 2022 evaluation study \textit{OVRseen} \cite{S22} indicated a significant lack of privacy measures in commercial-grade applications: $70$\% of VR data flows from the most widely adopted VR device were not appropriately disclosed, and $69$\% of them were used for purposes different to the core apps' functionality.
Hence, we encourage researchers to systematically open-source their code and engage with companies to bring privacy protections to consumers.

On the industry side, several indicators are forming a trend not conducive to enhancing privacy.
Companies have not shipped a ``VR incognito mode'' to avoid user tracking across VR applications, some developers ignore their own privacy policies \cite{S22}, consumers need to pay extra to sign in to their VR headsets without a social media account \cite{oculus_fb_signin}, the patents of a major VR company reveal how face tracking will help with personalized advertising in future metaverse applications \cite{meta_patents}, policy updates trend towards more data collection \cite{S22}, VR devices and applications are shown to be vulnerable to exploits \cite{S21}, and companies can ban plugins \cite{vrchat_mods} that could help with privacy, safety, and user disabilities.

Nonetheless, the most advanced commercial-grade VR device released on late October 2022 \cite{meta_quest_pro} has proposed a set of privacy features that update the industry standards \cite{meta_quest_privacy}: (i) monitoring features are turned off by default, (ii) tracking is paused at headset removal, (iii) cameras and microphone are turned off during headset's sleep mode, (iv) raw images are processed, stored, and deleted locally, (v) the extracted features are not used to identify users, and (vi) external lights on the headset signal bystanders that outwards cameras might record them.
However, there exist caveats, e.g., the company's eye-tracking notice \cite{eye_tracking_meta_notice} indicate that abstracted facial tracking information could be stored and processed by servers (e.g., potentially for psychographic profiling), abstracted gaze data can be shared with third parties (where the data is subject to their own privacy policies), and dark patterns are prevalent, namely, ``\textit{Enable}'' buttons are more highlighted than the ``\textit{Not Now}'' option.
Overall, given these industry and research privacy gaps, there are numerous opportunities for academics and practitioners to improve the state-of-the-art in privacy-enhancing VR systems, e.g., adding features such as end-to-end encryption for biometrics like eye-tracking, control programs to verify third party apps, more granular user privacy options, and abstracting biometric time-series to events \cite{VR_blog}.

\begin{table}[htpb!]
\caption{Min-Set coverage of attacks and defenses.}
\begin{center}
\vspace{-4mm}
{\footnotesize
\begin{tabular}{|p{12mm}||p{28mm}|p{28mm}|}
\hline
\textbf{Class} & \textbf{Privacy Attack} & \textbf{Privacy Defense} \\ \hline\hline
  
\multicolumn{3}{|l|}{\textbf{\textit{Identification}}}
\\ \hline  
 
\textbf{Geospatial} & (A1) \textit{MetaData} & \multicolumn{1}{l|}{(D1) \textit{MetaGuard} } 
\\ \cline{2-3}

\textbf{Telemetry} & (A12) \textit{BioMove}\textsuperscript{$\dagger$} & \multicolumn{1}{c|}{$\star$}  \\ \hline

\textbf{Inertial}	&	(A6) \textit{GaitLock}\textsuperscript{$\dagger$} &	\multicolumn{1}{c|}{$\star$}
\\ \cline{2-3}

\textbf{Telemetry} & (A5) \textit{Face-Mic} & \multicolumn{1}{c|}{$\star$} \\ \hline

\multirow{4}{*}{\textbf{Text}}	& (A28) Virtual Typing &	\multicolumn{1}{l|}{(D35) \textit{ReconViguRation}}
\\ \cline{2-3}

& \multirow{3}{*}{(A29) \textit{VR-Spy}} & \multicolumn{1}{l|}{(D35) \textit{ReconViguRation}\textsuperscript{$\star$},} 
\\ 

&  & \multicolumn{1}{l|}{VPN, Tor, Proxies,} 
\\ 

&  & \multicolumn{1}{l|}{End-To-End Encryption.} 
\\ \hline

\multirow{2}{*}{\textbf{Audio}}	& (A3) \textit{MitR} & \multicolumn{1}{l|}{(D33) \textit{MitR Defense}\textsuperscript{$\ddagger$}} 
\\ \cline{2-3} 

	& Speech Recognition	&	\multicolumn{1}{l|}{Voice Modulation \cite{nautsch_preserving_2019, S4}} 
\\ \hline 

\multirow{2}{*}{\textbf{Video}}	& (A13) Eye Tracking\textsuperscript{$\ddagger$}	 &	\multicolumn{1}{l|}{(D2) \textit{Kal$\varepsilon$ido}\textsuperscript{$\star$}}
\\ \cline{2-3}

& (A14) Iris Identifi.\textsuperscript{$\ddagger$}  & \multicolumn{1}{l|}{(D11) \textit{EyeVEIL}} 
\\ \hline

\textbf{Physio.} & 
\multicolumn{1}{c|}{\multirow{2}{*}{$\star$}} &
\multicolumn{1}{c|}{\multirow{2}{*}{$\star$}} 
\\  

\textbf{Signals} &  &   
\\ \hline

\textbf{System}	& (A1) \textit{MetaData}	& \multicolumn{1}{l|}{(D1) \textit{MetaGuard}\textsuperscript{$\star$}} 
\\ \hline 

\multirow{2}{*}{\textbf{Network}} & \multirow{2}{*}{(A1) \textit{MetaData}}	 &	\multicolumn{1}{l|}{(D1) \textit{MetaGuard}\textsuperscript{$\star$},} 
\\ 

&  & \multicolumn{1}{l|}{VPN, Tor, Proxies.} 
\\\hline\hline

\multicolumn{3}{|l|}{\textbf{\textit{Profiling}}}
\\ \hline

\textbf{Geospatial}	& (A1) \textit{MetaData} & \multicolumn{1}{l|}{(D1) \textit{MetaGuard}\textsuperscript{$\star$}} 
\\ \cline{2-3} 

\textbf{Telemetry} 	& (A10) Movement~Bio. &	\multicolumn{1}{c|}{$\star$} 
\\ \hline

\textbf{Inertial}	& (A5) \textit{Face-Mic}  &	 \multicolumn{1}{c|}{$\star$} 
\\ \cline{2-3}

\textbf{Telemetry} & (A10) Movement~Bio.   &\multicolumn{1}{c|}{$\star$}  \\ \hline

\textbf{Text}	&	\multicolumn{2}{l|}{\textit{Ditto}} 
\\ \hline

\multirow{2}{*}{\textbf{Audio}}	& (A1) \textit{MetaData} & \multicolumn{1}{l|}{(D1) \textit{MetaGuard}\textsuperscript{$\star$}} 
\\ \cline{2-3} 

	& (A3) \textit{MitR} &	\multicolumn{1}{l|}{(D33) \textit{MitR Defense}\textsuperscript{$\ddagger$}\textsuperscript{$\star$}} 
\\ \hline 

\multirow{3}{*}{\textbf{Video}}	&(A16) \textit{EMOShip}\textsuperscript{$\dagger$} & \multicolumn{1}{l|}{(D12) \textit{Kal$\varepsilon$ido}\textsuperscript{$\star$} }
\\ \cline{2-3} 

	& (A17) Spatial Recog.	&	\multicolumn{1}{l|}{(D16) Spatial Re.}
\\  \cline{2-3} 

	& (A18) Spatial Recog.\textsuperscript{$\ddagger$} & \multicolumn{1}{l|}{(D13) Spatial Re.}
\\  \hline

\textbf{Physio.}	& (A20) \textit{Vreed}\textsuperscript{$\dagger$} &\multicolumn{1}{c|}{$\star$} 
\\ \cline{2-3} 

\textbf{Signals}	& (A22) Signal~Proces.\textsuperscript{$\dagger$}	&	\multicolumn{1}{c|}{$\star$}
\\ \hline 

\multirow{3}{*}{\textbf{Behavior}}	& (A1) \textit{MetaData} & \multicolumn{1}{l|}{(D1) \textit{MetaGuard}\textsuperscript{$\star$} } 
\\ \cline{2-3} 

	& (A3) \textit{MitR} 	&	\multicolumn{1}{l|}{(D33) \textit{MitR Defense}\textsuperscript{$\ddagger$}\textsuperscript{$\star$} } 
\\ \cline{2-3} 

	& (A2) \textit{Malicious Design}	&	\multicolumn{1}{l|}{(D32) Design Defense\textsuperscript{$\ddagger$}\textsuperscript{$\star$}} 
\\ \hline 

\end{tabular}
}
\end{center}
{\small
\textsuperscript{$\dagger$}An attacker can leverage the associated defense/mechanism for adversarial purposes. \textsuperscript{$\ddagger$}While the study is attack/defense focused, there is a defensive/adversarial component. \textsuperscript{$\star$}Privacy opportunity.
}
\label{tab:min_set}
\vspace{-5mm}
\end{table}

\section{Discussion \& Future Work}
\label{sec:discussion}

We distill key findings (KF) and future work (FW) from studying the $74$ selected publications and our results:

\noindent \textbf{(KF1)} \textit{There is a fundamental imbalance in the deployment of offensive and defensive VR research.} 
While most reviewed papers focused on defensive techniques, none were deployed. This indicates limited knowledge transfer to industry, as evident by the absence of open-source code in academic papers.. In contrast, we have so far seen a plethora of device vulnerabilities \cite{S21}, an increased data collection and  privacy policy disregard \cite{S22}, along with one deployed academic offensive technique (A16). 
This contrasts with web privacy research, where vulnerabilities are highlighted yet countermeasures are widely available.


\noindent \textbf{(KF2)} \textit{Only a few defense studies ($17$\%) provided provable privacy guarantees.}
Provable privacy appeared in eye tracking studies (D2, D4-5, D8, D10) and spatial telemetry (D1).
However, provable privacy is still uncommon in VR, possibly due to its relative immaturity.


\noindent \textbf{(KF3)} \textit{VR authentication mechanisms should remain in the device.} 
Selected studies indicate biometric movement can leak identity, supporting the inclusion of continuous authentication within the device, such as secure enclaves akin to face ID verification on mobile phones, to prevent user switching during activities like exams.

\noindent \textbf{(KF4)} \textit{There is a lack of hardware-level privacy defenses.}
E.g., practitioners could use (D1) \textit{MetaGuard} at the firmware level or execute a signed function in a trusted execution partition instead of relying on performance variations to detect application misbehavior (D13).

\noindent \textbf{(KF5)} \textit{Attack benchmarks should use appropriate metrics, and defense proposals should generally include usability and performance studies.}
We encourage researchers to add \textit{F1-Scores} or \textit{equal error rates}, as false positives and negatives are essential for security and privacy, and measure performance degradation in, e.g., execution time, battery consumption (D14, D25), and usability (D1, D9).

\noindent \textbf{(KF6)} \textit{We identified the most dangerous attacks.}
Based on the literature, (A31) \cite{nair2023unique} and (A1) \textit{MetaData} are the most effective and practical identification and (broadest) profiling attacks, respectively. 


Beyond the privacy opportunities highlighted in \S\ref{sec:opportunities} and the limitations of RQ9, we encourage researchers to:

\noindent \textbf{(FW1)} \textit{Design protections for spatial and inertial telemetry against adversarial interference.}
The majority of surveyed defenses focused on protecting information extracted from video feeds.
Moving forward, we hope to see defensive research fill the remaining gap.

\noindent \textbf{(FW2)} \textit{Explore attacks of VR-native stimuli.} While some reviews discussed the safety dangers of maliciously manipulating video output \cite{VR_safety_manipulation, safety_II, VR_saftey_I}, none investigated if such dangers apply to audio, stereoscopic vision, haptic feedback, or other VR-specific outputs. There is a corresponding lack of defenses against such risks.

\noindent \textbf{(FW3)} \textit{Develop concrete countermeasures against malicious content design.} Mitigating the ability of adversaries to gain information or influence users by manipulating the immersive VR environment is amongst the most difficult open problems in VR. Achieving an appropriate balance between flexibility and consumer protection in VR environment design remains a significant outstanding challenge.

\noindent \textbf{(FW4)} \textit{Resume research in privacy protections for VR user interaction.}
The deeply immersive nature of VR makes social engineering a salient threat, but many extant studies on this subject are outdated by more than 10 years \cite{S1}.

\noindent \textbf{(FW5)} \textit{Study the inference of health conditions based on physiological signals in VR.}
The studies inferred emotions and arousal in VR, revealing how adversaries could deduce  neurological and physical disabilities, addictions, and health conditions like asthma, highlighting the deeply personal nature of VR privacy threats and need for defensive research.

\noindent \textbf{(FW6)} \textit{Explore the use of trusted execution environments (TEEs) in VR.}
At server or client level, TEEs could enhance privacy beyond surveyed defenses. Deploying TEEs for GPUs might open new privacy avenues.

\vspace{-2mm}
\section{Conclusion}
\label{sec:Conclusion}

In this SoK, we present a threat and defense framework for data privacy in VR and outline privacy opportunities for practitioners. Despite more defense proposals than attacks in literature, existing defenses are not exhaustive, some are missing, and most remain undeployed. The rise of data-hungry companies and pervasive data collection in VR highlights the need for increased cross-collaboration between industry and academia. Our frameworks and taxonomies aim to provide a foundation for future collaborations and research on metaverse privacy issues.

\bibliographystyle{ACM-Reference-Format}
\bibliography{999_References}

\appendix
\section{Detailed Data Collection Method \& Results}
\label{app_sec:method}

We inspired our method from seminal SoKs and reviews in the field of security and privacy \cite{SOK_Auth, PETs_IoT, S1}. For this SoK, we sought literature presenting at least one of the following artifacts in the context of virtual reality (VR) and privacy: (i) privacy threat or (ii) defense models, (iii) taxonomy of attributes or (iv) applications, (v) survey or implementation of attacks or (vi) defenses. 
When a paper encompassed mixed reality, we included the work if the presented artifacts (partially) overlapped with VR. 

Before the search, the researchers knew of $12$ relevant studies containing the target artifacts (base literature).
Two researchers curated the search string by studying the base literature and conducting a manual preliminary search in Google Scholar for papers containing the targeted artifacts.

\noindent \textbf{Search string}: (``virtual reality'' {\tt OR} ``virtual telepresence''  {\tt OR} ``head-mounted displays''  {\tt OR} ``head-worn display''  {\tt OR} ``metaverse'')
{\tt AND} ``data''  {\tt AND} ``privacy''  {\tt AND} (``attack''  {\tt OR} ``offense'' {\tt OR} ``defense''  {\tt OR} ``protection'') 
\smallskip

With this search string, we queried on August 10\textsuperscript{th}, 2022 the seven most relevant digital libraries: IEEE Xplore, ACM Digital Library, ScienceDirect, SpringerLink, Scopus, Wiley InterScience, and Web of Science.
We included work published in or after 2010 and excluded books, resulting in $1700$ hits. 
Two researchers filtered in parallel the publications by title ($47$ selected from $1700$), abstract ($35$ selected from $47$), and full text ($14$ selected from $35$), resolving conflicts in an informed discussion attending to the target artifacts and removing duplicates.
To reduce the number of possible missing publications, we employed Google Scholar to rerun the same search string and process.
We stopped when the publication titles did not contain keywords and added two new publications to the selected studies.
Combining the base literature ($12$) with the filtered studies ($16$) resulted in $23$ selected studies after deduplication. 

To further ensure we collected as many relevant publications as possible, we conducted a backward search of the references of the $23$ selected studies under the same criteria.
Additionally, we contacted the authors of the $23$ works to obtain further relevant publications.
The backward search revealed $26$ studies, and from the corpus signaled by the scholars, we included another $7$ after deduplication.
Lastly, we collected another $18$ publications thereafter throughout the research and writing of this manuscript following the same criteria (with publications as recent as March 2023).
Table~\ref{tab:SoK_studies} shows our final list of $74$ publications, and which of them are not peer-reviewed (only $4$).

\begin{table} [htpb!]

\caption{The $74$ collected studies. } 
\vspace{-6mm}

\small
\begin{center}
\begin{tabular}{|p{2.1cm}|p{5.75cm}|}
\hline

\textbf{Study focus} & \textbf{Studies ($74$)}
\\
\hline \hline 

\multicolumn{2}{|l|}{\textit{\textbf{VR Primary Studies} ($62$)}}
\\
\hline 
\textbf{Defenses} ($35$) & \cite{S5}, \cite{S7}, \cite{S8}, \cite{S9}, \cite{S11}\textsuperscript{$\dagger$}, \cite{S16}, \cite{S18}\textsuperscript{$\ddagger$}, \cite{S19}, \cite{S23}\textsuperscript{$\ddagger$}, \cite{S25}, \cite{S26}, \cite{S27}, \cite{S28}, \cite{S29}, \cite{S31}, \cite{S32}\textsuperscript{$\ddagger$}, \cite{S34}\textsuperscript{$\ddagger$}, \cite{S36}, \cite{S35}, \cite{S37}, \cite{S38}, \cite{S39}, \cite{S40}, \cite{S41}, \cite{S42}, \cite{S43}, \cite{S44}, \cite{S45}, \cite{S47}\textsuperscript{$\ddagger$}, \cite{S51}\textsuperscript{$\dagger$}, \cite{S53}\textsuperscript{$\dagger$}, \cite{S54}\textsuperscript{$\dagger$}, \cite{S55}\textsuperscript{$\dagger$}, \cite{S56}, \cite{S68}
\\\hline
\textbf{Attacks} ($23$) & \cite{S2}, \cite{S3}\textsuperscript{$\dagger$}, \cite{S13}\textsuperscript{$\ddagger$}, \cite{S15}, \cite{S30}, \cite{S46}, \cite{S49}\textsuperscript{$\dagger$}, \cite{S50}\textsuperscript{$\ddagger$}, \cite{S52}, \cite{S57}\textsuperscript{$\dagger$}, \cite{S58}\textsuperscript{$\dagger$}, \cite{S59}\textsuperscript{$\dagger$}, \cite{S60}\textsuperscript{$\dagger$}, \cite{S61}\textsuperscript{$\dagger$}, \cite{S62}\textsuperscript{$\dagger$}, \cite{S63}\textsuperscript{$\dagger$}, \cite{S64}\textsuperscript{$\dagger$}, \cite{S66}, \cite{S67}, \cite{nair2023unique}, \cite{N4}, \cite{N5}, \cite{N6}
\\\hline
\textbf{Surveys} ($2$) & \cite{S4}, \cite{S6}\textsuperscript{$\dagger$}
\\\hline
\textbf{Evaluations} ($2$) & \cite{S21}, \cite{S22}
\\\hline\hline 

\multicolumn{2}{|l|}{\textit{\textbf{VR Secondary Studies} ($12$)}} 
\\ \hline
\begin{tabular}[l]{@{}l@{}} \textbf{Literature} \\ \textbf{Reviews}\end{tabular} & \begin{tabular}[l]{@{}l@{}}  \cite{S1}, \cite{S10}, \cite{S12}, \cite{S14}, \cite{S17}, \cite{S20}, \cite{S24}, \\ \cite{S33}, \cite{S48}, \cite{S65}, \cite{N3}, \cite{N7} \end{tabular}
\\\hline
\end{tabular}
\end{center}

\begin{center}
\begin{tabular}{|l|}
\hline
\textit{\textbf{Selected Studies Not Peer-Reviewed} ($5\%$)}: \cite{S15}, \cite{S16}, \cite{S29}, \cite{S48} 
\\\hline
\end{tabular}
\end{center}

{\small
\textsuperscript{$\dagger$}An attacker can leverage the associated defense/mechanism for adversarial purposes. \textsuperscript{$\ddagger$}Although the study is defense/attack focused, there is an adversarial/defensive component. 
}
\label{tab:SoK_studies}
\end{table}

\end{document}